\documentclass{acmtrans2f}
\newtheorem{theorem}{Theorem}[section]

\newtheorem{proposition}[theorem]{Proposition}
\newtheorem{lemma}[theorem]{Lemma}
\newdef{definition}[theorem]{Definition}
\newdef{example}[theorem]{Example}
\newdef{remark}[theorem]{Remark}
\newdef{Notation}[theorem]{Notation}

\newcommand{\Ruled}[2]{\mbox{$\displaystyle\frac{#1}{#2}$}}

\newcommand{\siitv}{\type}
\newcommand{\tid}{{\cal I}}
\def\:{{:}}
\def\l{{\lambda}}
\def\ll{\l\!\!\!\!\;\l}
\def\btab{\begin{tabular}}
\def\etab{\end{tabular}}
\newcommand\wit[1]{\rule{#1mm}{0mm}}
\newcommand\hoog[1]{}

\newcommand{\md}{}
\newcommand{\forget}[1]{}
\def\bar{\begin{array}}
\def\ear{\end{array}}
\newcommand{\qedf}{}
\newcommand{\G}{\Gamma}

\newcommand{\I}{{\bf I}}
\newcommand{\N}{{\sf N}}
\newcommand{\FV}{\mbox{\it FV}}
\newcommand{\set}[1]{\{#1\}}
\newcommand{\bcd}{\mbox{\footnotesize${\cal BCD}$}}
\newcommand{\abo}{\mbox{\footnotesize${\cal AO}$}}

\newcommand{\bake}{\mbox{\footnotesize${\cal B}${\it a}}}
\newcommand{\ehr}{\mbox{\footnotesize${\cal EHR}$}}

\newcommand{\ffun}{{\sf fun}}

\newcommand{\into} {\cap}
\newcommand{\ti}{\lambda\cap}
\newcommand{\tin}{{\ti}_{{\cal B}}}
\newcommand{\tityt}{{\ti}_{\tyt}}
\newcommand{\titnu}{{\ti}_{\tnu}}

\newcommand{\con}{{\sf C}}
\newcommand{\conv}{\con^{\tytv}}

\newcommand{\arr} {\rightarrow}
\newcommand{\tyt} {\Omega}
\newcommand{\tnu} {\nu}

\newcommand{\typetytv}{\type^{\tytv}}
\newcommand{\type}{{\sf T}}

\newcommand{\tva} {A}
\newcommand{\tvb} {B}
\newcommand{\tvc} {C}

\newcommand{\tvav} {\psi}
\newcommand{\tvd} {D}
\newcommand{\tve} {E}

\newcommand{\sar}[1]{\Sigma^{#1}}
\newcommand{\tleq} {\leq}
\newcommand{\tleqt}{\tleq_{\tytv}}

\newcommand{\nottleqt}{\mbox{$\not\tleq$}_\tytv}

\newcommand{\binto} {\bigcap}

\newcommand{\teqt}{{}{\sim_{\tytv}}}
\newcommand{\notteqt}{{}{\mbox{$\not\sim$}_{\tytv}}}

\newcommand{\tytv} {\bigtriangledown}
\newcommand{\calT} {\sar{\tytv}}

\newcommand{\ctas}{\vdash^\tytv}
\newcommand{\ctasn}{\vdash^\tytv_{{\cal B}}}
\newcommand{\ctastyt}{\vdash^\tytv_\tyt}
\newcommand{\ctastnu}{\vdash^\tytv_\tnu}

\newcommand{\ax} {\mbox{Ax}}
\newcommand{\axtyt} {\mbox{Ax-$\tyt$}}
\newcommand{\axtnu}{\mbox{Ax-$\tnu$}}
\newcommand{\arrI} {\arr  \mbox{\rm I}}

\newcommand{\intI} {\into \mbox{\rm I}}
\newcommand{\arE} {\arr  \mbox{\rm E}}

\newcommand{\lleq} {\tleqt\mbox{\rm L}}

\newcommand{\intE} {\into \mbox{\rm E}}

\newcommand{\FX} {X}
\newcommand{\FY} {Y}

\newcommand{\SF}{{\cal F}^\tytv}

\newcommand{\FFF}{\uparrow}
\newcommand{\D} {{\cal D}}

\newcommand{\step}[2] {#1\Rightarrow#2}

\newcommand{\Dgeq} {\sqsupseteq}
\newcommand{\Dsup} {\sqcup}
\newcommand{\Dinf} {\sqcap}

\newcommand{\DSup} {\bigsqcup}

\newcommand{\env} {{\sf Env}}

\newcommand{\en} {\rho}
\newcommand{\ag} {\models}
\newcommand{\dlsqb}{[\![}
\newcommand{\drsqb}{]\!]}
\newcommand{\interpretation}[3]{\dlsqb{#1}\drsqb^{#3}_{#2}}

\newcommand{\ints}[2]{\dlsqb{#1}\drsqb^{\tytv}_{#2}}

\newcommand{\fun}{F^\tytv}

\newcommand{\gr}{G^\tytv}

\newcommand{\three}[3]{\langle {#1},{#2},{#3} \rangle}
\newcommand{\two}[2]{\langle {#1},{#2} \rangle}

\newcommand{\ten} {{\cal V}}

\newcommand{\agv} {\ag^{\tytv}}
\newcommand{\coni} {\con_\infty}

\newcommand{\four}[4] {\langle {#1},{#2},{#3},{#4} \rangle}

\firstfoot{ACM Transactions on Computational Logic,
                       Vol.\ TBD, No.\ TBD, TBD TBD, Pages \pages.}

            \runningfoot{ACM Transactions on Computational Logic,
                         Vol.\ TBD, No.\ TBD, TBD TBD.}
 \title{A Complete Characterization of Complete Intersection-Type Theories}
            \author{M. DEZANI-CIANCAGLINI\\
 Universit\`a di Torino \and
F. HONSELL and F. ALESSI\\ Universit\`a di Udine}

\begin{abstract}
We characterize those {\em intersection-type theories} which yield
complete {\em intersection-type assignment systems} for
$\lambda$-calculi, with respect to the three canonical set-theoretical
semantics for intersection-types: the inference semantics, the simple
semantics and the F-semantics. These semantics  arise
by taking as interpretation of types subsets of applicative structures, as interpretation of the {\em intersection constructor},
$\bigcap$, set-theoretic inclusion, and by taking the interpretation
of the {\em arrow constructor}, $ \rightarrow$, {\em \`a la } Scott, with respect to either {\em any possible functionality set}, or the {\em largest} one, or
the {\em least} one.

These results strengthen and generalize significantly all earlier
results in the literature, to our knowledge, in at least three
respects. First of all the  inference semantics had not been
considered before. Secondly, the characterizations are all given just
in terms of simple closure conditions on the {\em preorder relation} ,
$\leq$, on the types, rather than on the typing judgments
themselves. The task of checking the condition is made therefore
considerably more tractable. Lastly, we do not restrict attention just
to $\lambda $-models, but to arbitrary applicative structures which
admit an interpretation function. Thus we allow also for the treatment of models of
restricted $\lambda$-calculi. Nevertheless the characterizations we give can
be tailored just to the  case of $\lambda$-models.

\end{abstract}

            
\begin{document}
            \begin{bottomstuff}
            Author addresses: M. Dezani-Ciancaglini, Dipartimento di Informatica, Universit\`a di Torino, Corso
Svizzera 185, 10149 Torino, Italy  {\tt dezani@di.unito.it}. F. Honsell and F.Alessi, Dipartimento di Matematica ed
Informatica, Universit\`a di Udine, Via delle Scienze 208, 33100 Udine, Italy  {\tt honsell,
alessi@dimi.uniud.it}\\
Partially supported by MURST Cofin '99 TOSCA Project, FGV
'99 and CNR-GNSAGA.
            \permission{TBD}{TBD}
            \end{bottomstuff}
            \maketitle

\section{Introduction}

Intersection-types disciplines originated in \cite{coppdeza80} to overcome the
limitations of Curry's type assignment system and to provide a characterization of {\em
strongly normalizing terms} of the $\lambda$-calculus. But very early on, the
issue of {\em completeness} became crucial. Intersection-type theories and filter
$\lambda$-models have been introduced, in
\cite{barecoppdeza83}, precisely to achieve the
completeness for the type assignment
system $\ti^{\bcd}_\tyt$, with respect to Scott's simple semantics.  And this
result, together with the conservativity of $\ti^{\bcd}_{\tyt}$,
with respect to Curry's simple types, was used in~\cite{barecoppdeza83}
to prove Scott's conjecture concerning the completeness
of the set-theoretic semantics for simple types.

The number of type theories of interest in the literature has grown considerably over the
years   (e.g.
\cite{coppdezahonslong84,coppdezazacc87,honsronc92,egidhonsronc92,abraong93,plot93,honsleni99}, etc.), especially in connection with the study of domain models for $\lambda$-calculi in
the perspective of Abramsky's ``domain theory in logical form'' \cite{abra91}. Furthermore new semantics have been
proposed for intersection-types \cite{hind83}.

The problem of characterizing syntactically the sound and adequate (complete)
intersection-type theories,  with respect to the various set-theoretic semantics, appears therefore rather
natural. Moreover, we feel that the very existence of completeness results with respect to set-theoretic semantics, such as the one in
\cite{barecoppdeza83}, is probably one of the most significant features of
intersection-types.

In this paper we solve completely the characterization  problem as far as the three canonical
{\em set-theoretical} semantics for intersection-types: the inference semantics, the simple
semantics \cite{scot75} and the F-semantics \cite{scot80l}. These are the semantics which
arise by interpreting types as subsets of applicative structures, and by taking as
interpretation of the {\em intersection constructor},
$\cap$, set-theoretic inclusion, and by taking the interpretation
of the {\em arrow constructor}, $ \rightarrow$, {\em \`a la Scott} as a {\em logical predicate}, with respect  to either
{\em any possible functionality set}, or the {\em largest} one, or the {\em least} one.

More precisely, the {\em simple semantics}
of types associates to each arrow type $A\to B$ the set of
elements which applied to an arbitrary element in the interpretation of
$A$ return an element in the interpretation of $B$.

As Scott has pointed out in~\cite{scot80l}, however,  the key to a lambda model is the
set of elements in the domain which are {\em canonical representatives}
of functions, i.e. the elements which are meanings of terms starting
with an initial abstraction. The {F}-semantics of types takes therefore as meaning
of an arrow type only those elements which behave as expected with respect to
application {\em and} which are also canonical representatives
of functions.

The inference semantics is the counterpart of the inference semantics for polymorphic types
introduced in \cite{mitc88}, generalized to suitable applicative structures with an interpretation function, called $\lambda$-applicative structures. Here the
interpretation of arrows is taken with respect to an arbitrary set which includes the
canonical representatives of functions.

\newcommand{\Ic}{{\bf I}}
\newcommand{\Nc}{{\bf N}}

The results in this paper strengthen and generalize significantly all earlier
results in the literature, to our knowledge, in at least three
respects. First of all the  inference semantics had not been
considered before. Secondly the characterizations are all given just
in terms of simple closure conditions on the {\em preorder relation} ,
$\leq$, on the types, rather than on the typing judgments
themselves, as had been done earlier \cite{dezamarg86}. The task of
checking the condition is made therefore considerably more tractable. Lastly we do not
restrict attention just to $\lambda $-models, but to the more general class of $\lambda$-applicative structures. Thus we allow also for the treatment of models of
restricted $\lambda$-calculi, and most notably models of Plotkin's call-by-value
$\lambda_v$-calculus, and models of the $\lambda$-${\Ic}$-$\Nc$-calculus of \cite{honsleni99}.
Nevertheless the characterizations we give can be tailored just to the  case of
$\lambda$-models.

The paper is organized as follows.
In Section \ref{theories} we introduce
intersection-type theories, various kinds of
type assignment systems, and we prove Generation Lemmata for these systems.  In Section
\ref{appstruct} we introduce the basic semantical structures, with respect to which we shall
discuss soundness and completeness of intersection-type theories. In  Section
\ref{filter-structures-and-interpretation} we study filter structures and prove  the crucial
property satisfied by the  interpretation function over them.  Section
\ref{completeness-theorem} is the main section of the paper.  After introducing the notions
of  type interpretation domain and semantic satisfiability for the three semantics under
consideration, we give the characterization results.
Finally in Section \ref{fin} we discuss related results and give some final remarks.

\section{Intersection-type theories and type assignment systems}
\label{theories}
{\em Intersection-types} are syntactical objects which are built inductively by closing
a given set $\con$ of {\em type atoms} (constants) under the {\em function
  type} constructor $\arr $ and the {\em intersection} type
constructor $\into$.

\begin{definition}[Intersection-type Languages]
An {\em intersection-type language}, over $\con$, denoted by
  $\type=\type(\con)$ is defined by the following
abstract syntax:
 \[
\type = \con \mid \type \arr  \type \mid \type \into \type.\qedf
\]
\end{definition}

\begin{Notation}
  Upper case Roman letters i.e.\ $\tva,\tvb,\ldots$, will denote
  arbitrary types.  In writing intersection-types we shall use the
  following convention: the constructor $\into$ takes precedence over
  the constructor $\arr $ and it associates to the right.
Moreover $A^n\to B$ will be short for
$\underbrace{A\to\cdots\to A}_{n}\to B$.
\end{Notation}

 Much of the expressive power of intersection-type disciplines comes
 from the fact that types can be endowed with $\md$a$\md$ {\em
 preorder relation}, $\leq$, which induces the structure of a meet
 semi-lattice with respect to $\into$.  This appears natural
 especially in the semantical setting of the present paper, where the
 intended meaning of types are sets of denotations,
 $\into$ is interpreted as set-theoretic intersection,
 and $\leq$ is interpreted as set inclusion.

\begin{definition}[Intersection-type Preorder]\label{typetheory}
Let $\siitv=\type(\con)$ be an intersection-type language.
An {\em intersection-type preorder} over $\siitv$ is a binary relation
$\tleq$ on $\siitv$
satisfying the following set
$\tytv^0$ (``nabla-zero'')
of axioms and rules:
\[\begin{array}{ll}
\mbox{$\tva\tleq\tva$} & \mbox{(refl)}\\[0.5em]
\mbox{$\tva \tleq \tva \into \tva $} &
\mbox{(idem)}\\[0.5em]
\mbox{$\tva \into \tvb \tleq \tva $}&
\mbox{(incl$_L$)}\\[0.5em]
\mbox{$\tva \into \tvb\tleq \tvb $}&
\mbox{(incl$_R$)}\\[0.5em]
\Ruled{\tva \tleq \tva'\ \ \ \ \tvb \tleq \tvb' }{\tva \into \tvb
  \tleq \tva' \into \tvb'} & \mbox{(mon)}\\[1em]
  \Ruled{\tva \tleq \tvb \ \ \ \  \tvb \tleq \tvc}{\tva \tleq
  \tvc } &{\mbox{(trans)}}\qedf
\end{array}
\]

\end{definition}

\begin{Notation}
  We will write $\tva \sim \tvb$ for $\tva \tleq \tvb$ and $\tvb \tleq
  \tva$.
 \end{Notation}

\noindent Notice that associativity and commutativity of
$\tleq$ (modulo $\sim$) follow easily from the above axioms and
rules.

\begin{Notation}
Being $\into$ commutative and associative, we will write
  $\bigcap_{i \leq n} \tva_i$ for $\tva_1\into \ldots \into\tva_n$.
  Similarly we will write $\into_{i\in I}A_i$ where we convene that $I$
denotes always a finite non-empty set.
\end{Notation}

Possibly effective, syntactical presentations of
intersection-type preorders can be given using the notion of {\em
  intersection-type theory}. An intersection-type theory includes always the basic set $\tytv^0$ for $\tleq$ and possibly other of
special purpose axioms and rules.

\begin{definition}[Intersection-type Theories]\label{itt} Let
$\type=\siitv(\con)$ be an intersection-type language, and let $\tytv$
be a collection of axioms and rules for deriving judgments of the
shape $\tva\tleq\tvb$, with $A, B\in\type$.  The {\em intersection-type theory\/} $\Sigma(\con,{\tytv})$ is the set of all judgments
$\tva\tleq\tvb$ derivable from the axioms and rules in $\tytv^0\cup
\tytv$.$\qedf$
\end{definition}

\begin{Notation}
When we consider the intersection-type theory
$\Sigma(\con,{\tytv})$, we will write
$$\btab{lcl}
$\con^{\tytv}$& for & $\con$,\\
$\type^{\tytv}$& for & $\type(\con)$,\\
$\Sigma^{\tytv}$& for & $\Sigma(\con,{\tytv}).$
\etab$$
 Moreover $\tva\tleqt\tvb$ will be short for
  $(\tva\tleq\tvb)\in\sar{\tytv}$.  Finally we will write $A\teqt B\iff
  A\tleqt B\tleqt A$.
\end{Notation}

\begin{center}
\begin{figure}\label{f01}
$$\begin{array}{|llll|}
\hline
\hoog{2.5ex}
\wit{3}\mbox{($\tyt$)}\wit{9} &\tva \tleq \tyt&&\\[0.5em]
\wit{3}\mbox{($\tnu$)}\wit{9} &\tva\arr\tvb \tleq \tnu&&\\[0.5em]
\wit{3}\mbox{($\tyt$-$\eta$)}&\tyt \tleq \tyt \arr  \tyt&&\\[0.5em]
\wit{3}\mbox{($\tyt$-{\it lazy})} &
\tva \arr  \tvb \tleq \tyt \arr  \tyt&&\\[0.5em]
\wit{3}\mbox{($\arr $-$\into$)}\wit{9}& (A\arr B)\into(A\arr C)\leq A\arr B\into
 C&&\\[0.5em]
\wit{3}(\eta) &\Ruled{\tva' \tleq \tva\ \ \ \ \ \tvb \tleq \tvb'}{ \tva  \arr  \tvb
\tleq \tva' \arr  \tvb'}&&\\[1em]
\hline
\end{array}$$
\caption{Some special purpose Axioms and Rules concerning $\tleq$.}\label{f1}
\end{figure}
\end{center}

In Figure \ref{f1} appears a list of special purpose axioms and
rules which have been considered in the literature.
We give just a few lines  of motivation for each.

Axiom ($\tyt$)
states that the resulting type  preorder has a maximal element.
Axiom ($\tyt$) is particularly meaningful
when used in combination with the $\tyt$-type assignment
system, which essentially treats $\tyt$ as the
universal type of all $\lambda$-terms (see Definition
\ref{typeasstyt}).

The meaning of the other axioms and rules can be grasped easily if we consider
 again the intended set-theoretic semantics, whereby types denote subsets of
  a domain of discourse, and we interpret $\tva \arr \tvb$ as the
set of functions which map each element of $\tva$ into an
element of $\tvb$.

For instance, in combination with Axiom ($\tyt$), Axiom
($\tyt$-$\eta$) expresses the fact that all the objects in our domain
of discourse are total functions, i.e.\ that $\tyt$ is equal to $\tyt
\arr \tyt$~\cite{barecoppdeza83}.

However, if we want to capture only
those terms which truly represent functions, as it is necessary, for instance, in discussing the lazy $\lambda$-calculus ~\cite{abraong93}, we cannot assume axiom ($\tyt$-$\eta$) in order to ensure that all functions are total. To this end we can
postulate instead the weaker property ($\tyt$-{\it lazy}).
According to the set theoretic semantics, this axiom states, in effect,
simply that an element which is a function,
(since it maps
$\tva$ into $\tvb$) maps also the whole universe into itself.

The set-theoretic meaning of Axiom ($\arr $-$\into$) is immediate: if a function maps $\tva$ into $\tvb$, and  also $\tva$ into $\tvc$, then, actually, it maps the
whole $\tva$ into the intersectionof $\tvb$ and $\tvc$ (i.e.\
into $\tvb \into \tvc$), see \cite{barecoppdeza83}.

Rule $(\eta)$ is also very natural set-theoretically: it asserts
the arrow constructor is contravariant
in the first argument and covariant in the second one. Namely, if a function maps $\tva$ into $\tvb$, and we take a subset
$\tva'$ of $\tva$ and a superset $\tvb'$ of $\tvb$, then this function
will map also $\tva'$ into $\tvb'$, see~\cite{barecoppdeza83}.

Axiom $(\tnu)$ states that $\tnu$ includes any arrow type.
This axiom agrees with the $\tnu$-type assignment system, which
treats $\tnu$ as the universal type of all $\lambda$-abstractions
(see Definition \ref{typeasstnu}).
Notice that,  when the type denoting the whole
universe, $\tyt$ is in $\conv$, the role of $\tnu$ could be played also by the
type
$\tyt\arr\tyt$.
For this reason it is of no use to have at the same time in the language both
$\tnu$ and $\tyt$. Hence we impose that the two constants
do not occur together in any $\con^\tytv$.
The elements $\tyt$ and $\tnu$ play very special roles in the
development of the theory. Therefore we stipulate the following
blanket assumptions:\label{ba}
\[\begin{array}{ll}
 \mbox{{\bf Assumption 1}}: & \mbox{if $\tyt\in\con^\tytv$ then $(\tyt)\in\tytv$}.
\\
 \mbox{{\bf Assumption 2}}: & \mbox{if $\tnu\in\con^\tytv$ then $(\tnu)\in\tytv$}.
\end{array}
\]

We introduce in Figure \ref{f2} a list of significant intersection-type theories
which have been extensively considered in the literature. The order is
logical, rather than historical:
\cite{bake92,egidhonsronc92,abraong93,barecoppdeza83}.

We shall denote such theories as $\sar{\tytv}$, with various different
names $\tytv$ corresponding to the initials of the authors who have first considered the
$\lambda$-model induced by such a theory.  For each such $\tytv$ we
specify in Figure \ref{f2} the type theory
$\sar{\tytv}=\Sigma(\con,{\tytv})$ by giving the set of
constants $\con^{\tytv}$ and the set  $\tytv$ of extra axioms and rules
taken from Figure \ref{f1}. Here $\coni$ is an infinite set of fresh atoms, i.e.\ different from $\Omega,\nu$.
\renewcommand{\sar}[1]{{#1}}

\begin{center}
\begin{figure}\label{f2}
  $$ \begin{array}{|llllll|} \hline \hoog{3ex}
  \wit{1} \con^{\bake}&=&\coni&\bake&=& \{ \mbox{($\arr
      $-$\into$)}, (\eta)\}\\[0.5em]
  \wit{1} \con^{\ehr}&=&\{\nu\}&

  \ehr&=& \bake\cup\{(\nu)\}\\[0.5em]
  \wit{1} \con^{\abo}&=&\{\tyt\}&\sar{\abo}&=
  &\bake\cup\{(\tyt),\mbox{($\tyt$-{\it lazy})}\}\\[0.5em]
  \wit{1}\con^{\bcd}&=&\{\tyt\}\cup\coni\wit{3}\wit{1}&\sar{\bcd}&=&
  \bake\cup\{(\tyt),\mbox{($\tyt$-{$\eta$})}\} \\[0.5em]
\hline
\end{array}$$
\caption{Type Theories: atoms, axioms and rules.}\end{figure}
\end{center}

\renewcommand{\sar}[1]{\Sigma^{#1}}

Now that we have introduced intersection type theories we have to explain
how to capitalize effectively on their expressive power.
This is achieved via the crucial notion of {\em intersection type assignment system}.
This is a natural extension of Curry's type assignment type to intersection types.
First we need some preliminary definitions and notations.

\begin{definition}\label{definition-for-type-assignment}
\begin{enumerate}
\item
 A ${\tytv}$-{\em basis} is a set
  of statements of the shape $x\:\tvb$, where $\tvb\in\type^{\tytv}$,
  all whose variables are distinct.
\item  An {\em intersection-type
    assignment system}\ relative to $\sar{\tytv}$, denoted by
  ${\ti}^{\tytv}$, is a formal system for deriving judgments of the
  form $\Gamma \ctas M:\tva$, where
  the {\em subject\/} $M$ is an untyped $\lambda$-term,
  the {\em predicate\/} $\tva$ is  in $\type^{\tytv}$,
  and $\Gamma$ is a ${\tytv}$-basis.
\item  We will write $x\in\Gamma$ as short for
$\exists\tva.\;(x\:\tva)\in\Gamma$, i.e.\  $x$ occurs as the subject of an
assertion in $\Gamma$.
\item We say that a term $M$ is {\em typable} in
${\ti}^{\tytv}$, for a given ${\tytv}$-basis $\Gamma$, if there is a
type $A\in\type^\tytv$
such that the judgment $\Gamma \ctas M: A$ is derivable.
$\qedf$
\end{enumerate}
\end{definition}

\begin{definition}[Basic
    Type Assignment System]\label{typeassneutral}$\;$\\
 Let $\sar{\tytv}$ be a type theory.
 The {\em basic type assignment system},
denoted by $\tin^{\tytv}$, is a formal system for deriving judgments
of the shape $\Gamma \ctasn M:\tva$. Its rules are the following:

\[\begin{array}{lll}
 (\ax)&\Ruled{x\:\tva\in\Gamma}{\Gamma \ctasn x\:\tva} & \\[1em]
 (\arrI )&\Ruled{\Gamma, x\:\tva\ctasn M:\tvb}{\Gamma \ctasn \lambda
    x.M:\tva\arr \tvb} & \\[1em]
(\arE)&\Ruled{\Gamma \ctasn
    M:\tva\rightarrow\tvb\;\;\;\Gamma \ctasn N:\tva}
   {\Gamma \ctasn MN:\tvb}&\\[1em]
  (\intI)&\Ruled{\Gamma \ctasn M:\tva\;\;\;\Gamma \ctasn M:\tvb}{\Gamma \ctasn
    M:\tva\into\tvb}& \\[1em]
(\tleqt)&\Ruled {\Gamma \ctasn
    M:\tva\;\;\;\tva\tleqt\tvb}{\Gamma \ctasn M:\tvb}&\\[1em]
\end{array}\qedf\]
\end{definition}

The Basic Type Assignment System can be extended with other rules according to the
set of constants belonging to
$\conv$ and the corresponding axioms and rules in
$\sar{\tytv}$.

If
$\tyt\in\conv$, in line with the intended set-theoretic interpretation of $\tyt$ as
the universe, we extend the Basic Type Assignment System with a suitable axiom for
$\tyt$.

\begin{definition}[$\tyt$-type Assignment System]\label{typeasstyt}$\;$\\
  Let $\sar{\tytv}$ be a type theory
   with $\tyt\in\conv$.  The axioms and rules of the
  {\em $\tyt$-type assignment system\/} (denoted $\tityt^{\tytv}$), are those of
   the Basic type Assignment System, together with  the further axiom\\
\indent\indent
 $\begin{array}{lll}
(\axtyt)& \Gamma \ctastyt M:\tyt.\qedf
\end{array}$
\end{definition}

Similarly,
if $\tnu\in\conv$, in line with the intended interpretation of
$\tnu$  as the universe of {\em abstractions}, we define:

\begin{definition}[$\tnu$-type Assignment System]\label{typeasstnu}$\;$\\
  Let $\sar{\tytv}$ be a type theory with $\tnu\in\conv$.
   The axioms and rules of the
  {\em $\tnu$-type assignment system\/} (denoted $\titnu^{\tytv}$), are those of
   the Basic Type Assignment System, together with the further axiom\\
\indent\indent
 $\begin{array}{ll}
(\axtnu)& \Gamma \ctastnu \lambda x.M:\tnu.\qedf
\end{array}$
\end{definition}

For ease of notation, we convene that the symbols $\tyt$ and $\tnu$ are reserved for
the distinguished  type constants used in the systems $\tityt^{\tytv}$
and $\titnu^{\tytv}$, and hence  we forbid $\tyt\in\conv$ or $\tnu\in\conv$ when we
deal with $\tin^{\tytv}$.

\begin{Notation}In the following  ${\ti}^{\tytv}$ will
range over $\tin^{\tytv}$, $\tityt^{\tytv}$
and $\titnu^{\tytv}$. More precisely we assume that
${\ti}^{\tytv}$ stands for $\tityt^{\tytv}$
whenever $\tyt\in\conv$, for $\titnu^{\tytv}$
whenever $\tnu\in\conv$, and for $\tin^{\tytv}$ otherwise.
Similarly for $\ctas$.
\end{Notation}

We refer to \cite{bare00} for
a detailed account on the interest and differences of the three
intersection-type assignment systems introduced above.  Here we just recall a few suggestive facts. Thanks to the
intersection-type constructor,  {\em self-application}  can  be typed in the system
$\tin^{\tytv}$, while this was not the case  with Curry's type assignment system.
For instance it is easy to prove that
$\ctasn\lambda x.xx:(\tva\arr \tvb)\into\tva\arr \tvb$.
 Actually, all strongly normalizing terms are typeable
in $\tin^{\tytv}$.
All solvable terms can be typed in
  $\tityt^{\tytv}$ with some type not equivalent to $\tyt$.
 For instance, using axiom (Ax-$\tyt$),
the term $(\lambda yx.x)(\Delta\Delta)$, where
 $\Delta\equiv\lambda x.xx$,  can be given type
$\tva\arr\tva$.
The system $\titnu^{\tytv}$ is appropriate for dealing with Plotkin's
call-by-value $\lambda_v$-calculus. Also this system allows  to type
non-strongly normalizing terms. For instance, one can prove that the term
$(\lambda yx.x)(\lambda z.\Delta\Delta)$ may receive type
$\tva\arr\tva$ for all $\tva$. Anyway, notice that,
as proved in \cite{egidhonsronc92},
$(\lambda yx.x)(\Delta\Delta)$ cannot be typed in
$\titnu^{\tytv}$.

Notice that the structural rules of {\em (weakening)} and {\em
(strengthening)} are admissible in all ${\ti}^{\tytv}$¹s:
$$\begin{array}{rlrl}
(\mbox{\it weakening}) & \Ruled{\Gamma \ctas M:\tva}{\Gamma,x:\tvb \ctas M:\tva}
& (\mbox{\it strengthening}) &
\Ruled{\Gamma \ctas M:\tva}{\Gamma\lceil M \ctas M:\tvb},
\end{array}$$
where $\Gamma\lceil M =\{x:\tvb\; |\; x\in \FV(M)\}$.

Notice also that the intersection elimination rules
$$\begin{array}{llll}
(\intE) & \Ruled{\Gamma \ctas M:\tva\into\tvb}{\Gamma \ctas M:\tva}
&\ \ \ \ &
\Ruled{\Gamma \ctas M:\tva\into\tvb}{\Gamma \ctas M:\tvb}.
\end{array}$$
can be proved immediately to
be derivable in all ${\ti}^{\tytv}$'s.\\
Moreover, by a straightforward
induction on the structure of derivations, one can prove  that the rule
$$\begin{array}{ll}
(\lleq)&\Ruled{\Gamma, x\: \tvb \ctas M:\tva \;\;\; \tvc \tleqt  \tvb}
{\Gamma ,x\:\tvc \ctas M:\tva};\\[1em]
\end{array}$$
 is admissible in all ${\ti}^{\tytv}$'s.

\hbox to \hsize{\indent We conclude this section by proving a crucial technical result concerning}
\noindent intersection-type theories, which will be useful in Section \ref{completeness-theorem}. It is a form of
generation (or inversion) lemma, which provides conditions for  ``reversing'' some
of the rules of the type assignment systems
${\ti}^{\tytv}$.

\begin{Notation}
When we write ``...assume $\tva\notteqt\tyt$...'' we mean that
this condition is always true when we deal with
$\ctasn$ and $\ctastnu$, while it must be checked for
$\ctastyt$. Similarly, the condition $\tnu\nottleqt\tva$ must
be checked just for $\ctastnu$.

Moreover we write ``the type theory $\sar{\tytv}$ validates $\tytv'$''
to mean that all axioms and rules of
$\tytv'$ are admissible in $\sar{\tytv}$.
\end{Notation}

\begin{theorem}[Generation Lemma]\label{gen-l}
Let $\sar{\tytv}$ be a type theory.
\begin{enumerate}
\item \label{gen-l1}
Assume $ \tva\notteqt \tyt $. Then
$\Gamma\ctas x:\tva$ iff $(x\:\tvb)\in \Gamma$ and $\tvb\tleqt \tva$ for some
$\tvb \in \type^{\tytv}$.
\item \label{gen-l2}
Assume $ \tva\notteqt \tyt $. Then
$\Gamma\ctas MN:\tva$ iff $\Gamma\ctas M:\tvb_i\arr \tvc_i$,
$\Gamma\ctas N:\tvb_i$, and
$\binto_{i \in I}\tvc_i\tleqt \tva$ for some $I$
and $ \tvb_i, \tvc_i \in{\type}^{\tytv}$.
\item \label{gen-l4}
Assume $ \tva\notteqt \tyt $ and let $\sar{\tytv}$ validate $\bake$. Then
$\Gamma\ctas MN:\tva$ iff $\Gamma\ctas M:\tvb\arr \tva$, and $\Gamma\ctas N:\tvb$ for some
$\tvb \in \type^{\tytv}$.
\item \label{gen-l5}
Assume $ \tnu\nottleqt \tva $. Then  $\Gamma\ctas \lambda x.M:\tva$ iff
$\Gamma,x\:\tvb_i\ctas M:\tvc_i$, and
$\binto_{i \in I}(\tvb_i\arr \tvc_i)\tleqt \tva$ for some $I$
and $\tvb_i, \tvc_i \in {\type}^{\tytv}$.
\end{enumerate}
\end{theorem}

\begin{proof} The proof of each ($\Leftarrow$) is easy.
So we only treat ($\Rightarrow$).

(\ref{gen-l1}) Easy by induction on
derivations, since only the axioms ($\ax$), $(\axtyt)$, and the rules
$(\intI)$, $(\tleqt)$ can be applied. Notice that the condition
$ \tva\notteqt \tyt $ implies that $\Gamma\ctas x:A$
cannot be obtained just using axioms (Ax-$\tyt$).

(\ref{gen-l2}) By induction on derivations. The only interesting case
is when $A\equiv A_1\into A_2$ and the last rule  applied is ($\intI$):
$$\begin{array}{ll}
(\intI)&\Ruled{\Gamma\ctas MN:\tva_1\;\;\;\Gamma
\ctas MN:\tva_2}{\Gamma\ctas MN:\tva_1\into A_2}.
\end{array}$$
The condition $ \tva\notteqt \tyt $ implies that we cannot have
$ \tva_1\teqt \tva_2\teqt \tyt $. We do the proof for $ \tva_1\notteqt \tyt $
and $ \tva_2\notteqt \tyt $, the other cases can be treated similarly.
By induction there are $I,B_i,C_i,J,D_j,E_j$ such that
$\forall i \in I.$ $\Gamma\ctas M:\tvb_i\arr \tvc_i$, $\Gamma\ctas N:\tvb_i$,
$\forall j \in J.$ $\Gamma\ctas M:\tvd_j\arr \tve_j$, $\Gamma\ctas N:\tvd_j$,
and moreover $\binto_{i \in I}\tvc_i\tleqt \tva_1$,
$\binto_{j \in J}\tve_j\tleqt \tva_2$.
 So we are done since
$(\binto_{i \in I}\tvc_i)\cap(\binto_{j \in J}\tve_j)\tleqt \tva$.

(\ref{gen-l4}) Let $I, \tvb_i, \tvc_i$ be as in (\ref{gen-l2}).
Applying rule  $(\intI)$ to $\Gamma\ctas M:\tvb_i\arr \tvc_i$ we can derive
$\Gamma\ctas M:\binto_{i \in I}(\tvb_i\arr \tvc_i)$, so by $(\tleqt)$ we have
$\Gamma\ctas M:\binto_{i\in I}\tvb_i\arr \binto_{i \in I}\tvc_i$,
since\\ $\binto_{i \in I}(\tvb_i\arr \tvc_i)\tleqt
\binto_{i \in I}(\binto_{i\in I}\tvb_i\arr \tvc_i)
\tleqt\binto_{i\in I}\tvb_i\arr \binto_{i \in I}\tvc_i$ by rule ($\eta$)
and axiom ($\arr$-$\into$).

We can choose $\tvb=\binto_{i\in I} \tvb_i$ and conclude
$\Gamma\ctas M:\tvb\arr \tva$ since $\binto_{i \in I}\tvc_i\tleqt \tva$.

(\ref{gen-l5}) If $ \tva\teqt \tyt $
we can choose $\tvb\equiv\tvc\equiv\tyt$.
Otherwise $ \tva\notteqt \tyt $ and $  \tnu\nottleqt \tva  $.
The proof is
by induction on derivations.
Notice that $\Gamma\ctas \lambda x.M:\tva$ cannot be
obtained just using axioms (Ax-$\tyt$) or (Ax-$\tnu$).
The only interesting case is again when $A\equiv A_1\into A_2$ and the last
 rule applied is ($\intI$):
$$\begin{array}{ll}
(\intI)&\Ruled{\Gamma\ctas\lambda x.M:\tva_1\;\;
\;\Gamma\ctas \lambda x.M:\tva_2}{\Gamma\ctas \lambda x.M:\tva_1\into A_2}.
\end{array}$$
As in the proof of (\ref{gen-l2}) we only consider the case
$ \tva_1\notteqt \tyt $, $\tnu\nottleqt A_1$,
$ \tva_2\notteqt \tyt $, and $\tnu\nottleqt A_2$.
By induction there are $I,B_i,C_i,J,D_j,E_j$ such that
$$\begin{array}{l}

\forall i\in I.\ \Gamma,x\:B_i\ctas M: \tvc_i,
\ \ \forall j\in J.\ \Gamma,x\:D_j\ctas M:\tve_j, \\
\binto_{i \in I}(\tvb_i\arr\tvc_i)\tleqt \tva_1\ \ \&\ \
\binto_{j \in J}(\tvd_j\arr\tve_j)\tleqt \tva_2.
\end{array}$$
So we are done since
$(\binto_{i \in I}(\tvb_i\arr\tvc_i))
\cap(\binto_{j \in J}(\tvd_j\arr\tve_j))\tleqt \tva$.$\qed$
\end{proof}

Special cases of this theorem have already appeared in the literature, see
\cite{barecoppdeza83,coppdezahonslong84,coppdezazacc87,honsronc92,egidhonsronc92}.


\section{Applicative structures suitable for lambda calculus}
\label{appstruct}

In this section we introduce the semantical structures which we will
consider in our investigation of soundness and completeness of
intersection-type assignment systems.

\begin{definition}[$\lambda$-applicative structure]
A {\em $\lambda$-applicative structure} is a triple
$\langle \D, \cdot, \interpretation{\;}{\;}{\D}\rangle$ such that:
\begin{enumerate}\item $\two{\D}{\cdot}$ is  an applicative structure;
\item $\interpretation{\;}{\;}{\D}:\Lambda\times{\sf Env}_\D\to\D$, where
${\sf Env}_\D= [ \mbox{\it Var} \rightarrow \D]$,
is an  interpretation function for $\lambda$-terms.
\end{enumerate}

\end{definition}

$\lambda$-applicative structures are those applicative structures, which
have just enough structure to interpret the language of
$\lambda$-calculus.

It is often the case that we want to focus on the class of
$\lambda$-applicative structures which provide a {\em compositional}
interpretation function.  Hence we introduce:

\begin{definition}[Compositional $\lambda$-applicative structure]
A {\em compositional $\lambda$-applicative structure} is a
$\lambda$-applicative structure, where the interpretation function
satisfies the following properties
\begin{enumerate}
\item $\interpretation{MN}{\rho}{\D}  =
\interpretation{M}{\rho}{\D} \cdot \interpretation{N}{\rho}{\D}$;
\item $\interpretation{\lambda x.M}{\rho}{\D}
= \interpretation{\lambda y.M[x:=y]}{\rho}{\D}$ if $y\not
\in \FV(M)$;
\item $(\forall d\in \D.\  \interpretation{M}{\rho[x:=d]}{\D}  =
\interpretation{N}{\rho[x:=d]}{\D})
\Rightarrow \interpretation{\lambda x.M}{\rho}{\D} =
\interpretation{\lambda x.N}{\rho}{\D}$.
\end{enumerate}
\end{definition}

One can easily see that Plotkin's $\lambda$-structures
as defined in \cite{plot93},  are compositional $\lambda$-applicative structures.
In the next section we will introduce filter structures, which are again compositional $\lambda$-applicative structures.

Models of, possibly restricted,  $\lambda$-calculi as we commonly know them, can be viewed as special compositional $\lambda$-applicative structures.

First we need to give the definition of {\em restricted}  $\lambda$-calculus.

\begin{definition}[Restricted $\lambda$-calculus]
Let $R\subseteq \{ \langle (\lambda x.M)N, M[x:=N] \rangle \mid M,N \in \Lambda\}$. The
{\em  restricted $\lambda$-calculus} $\lambda_R$, is the  calculus
obtained from the standard $\lambda$-calculus, by restricting the
$\beta$ rule to the redexes in $R$.
\end{definition}

Clearly when $R=\beta$, $\lambda_R$ is the standard $\lambda$-calculus.
The main examples of {\em truly} restricted $\lambda$-calculi are Plotkin's call-by-value $\lambda_v$-calculus and
the $\lambda$-$\I$-$\N$-calculus of \cite{honsleni99}. Finally we give the crucial definition

\begin{definition}[(Restricted) compositional $\lambda$-model]
A {\em (restricted) compositional $\lambda$-model} for the (restricted)
$\lambda$-calculus $\lambda_R$, is a compositional $\lambda$-
applicative structure, $\langle \D, \cdot,
\interpretation{\;}{\;}{\D}\rangle$, which moreover satisfies
\[\interpretation{(\lambda x.M)N}{\rho}{\D} =  \interpretation{M[x:=N]}{\rho}{\D} \mbox{ for }  \langle (\lambda x.M)N, M[x:=N] \rangle \in R.\]
\end{definition}

It is easy to see that all notions of models for, possibly  restricted, $\lambda$-calculi,  based on applicative structures,
 can be cast in the above setting.


\section{Filter structures and interpretation of
lambda terms}\label{filter-structures-and-interpretation}

In this section we introduce {\em filter structures}. These are the basic tool for building $\lambda$-applicative structures, in effect $\lambda$-models, which realize completeness for intersection-type theories.

Filter structures arise naturally in the context of those generalizations of Stone
duality that are used in discussing domain theory in logical form (see
\cite{abra91}, \cite{coppdezahonslong84}, \cite{vick89}).

This approach provides, a conceptually independent semantics to intersection-types, the {\em lattice semantics}. Types are viewed as
{\em compact elements} of domains. The type $\Omega$
denotes the least element, intersections denote joins of compact
elements, and arrow types allow to internalize the space of continuous
endomorphisms.  Following the paradigm of Stone duality, type theories give
rise to {\em filter structures\/}, where the interpretation of
$\lambda$-terms can be given through a finitary logical description.

We start by introducing the  notion of {\em filter\/} of types.
Then we show how to associate to each type theory its
filter structure. This is a compositional $\lambda$-applicative structure where the  interpretation
of a $\lambda$-term is given by the filter of the types which can be assigned to it.

\begin{definition}\label{filter} Let $\calT$
be a type theory.
\begin{enumerate}
\item A $\tytv$-filter (or a filter over $\typetytv$) is a set
$\FX\subseteq \typetytv$ such that
\begin{enumerate}
\item
if $\tyt \in \con^{\tytv}$ then $\tyt \in \FX$;
\item
if $\tva \tleqt \tvb$ and $\tva \in \FX$, then $\tvb \in \FX$;
\item
if $\tva,\tvb \in \FX$, then $\tva\into\tvb \in \FX$;
\end{enumerate}
\item $\SF$ denotes the set of $\tytv$-filters over $\typetytv$;
\item if $\FX\subseteq \typetytv$,
$\FFF\FX$ denotes the $\tytv$-filter generated by $\FX$;
\item a $\tytv$-filter is {\em principal} if it is of the shape $\FFF
  \{ \tva\}$, for some type $\tva$. We shall denote $\FFF \{ \tva\}$
  simply by $\FFF \tva$. $\qed$
\end{enumerate}
\end{definition}

\noindent Notice that $\mbox{$\FFF\emptyset$ is the $\tytv$-filter $\FFF\tyt$,
  if $\tyt \in \con^{\tytv}$, and $\emptyset$}$ otherwise.

It is not difficult to prove that
$\SF$, ordered by subset inclusion,
is an $\omega$-algebraic complete lattice,
whose bottom element is $\FFF\emptyset$, and whose top element is
$\typetytv$. Moreover if $\FX,\FY\in\SF$,
$\FX\Dsup\FY=\FFF(\FX\cup\FY)$,
$\FX\Dinf\FY =\FX\into\FY$.
The sup of a directed set of filters is the set-theoretic union of filters.
The finite elements are exactly the filters generated by finite sets of types.

The next step is to define
application over sets of filters.

\begin{definition}[Filter structure]\label{application}
Let $\calT$ be a type theory.
\begin{enumerate}
\item \label{application1}
Application $\cdot:\SF\times\SF\rightarrow\SF$  is defined as
\[ X\cdot Y =\FFF\{ \tvb\mid \exists \tva\in \FY.\tva\to\tvb\in \FX\}.\]
\item \label{application2}
The maps $\fun:\SF\arr [\SF\arr \SF]$ and
$\gr:[\SF\arr \SF]\arr \SF$ are defined by:
\[\begin{array}{lll}\fun(\FX) &=& \lambda\!\!\lambda\FY\in\SF.\FX\cdot \FY;\\
\gr(f) &=&\left\{ \begin{array}{ll}
\FFF \{\tva\arr \tvb\mid\tvb\in f(\FFF\tva)\}\cup \uparrow \nu&
\mbox{if } \nu \in
\con^\tytv\\
\FFF \{\tva\arr \tvb\mid\tvb\in f(\FFF\tva)\}& \mbox{ otherwise.}
\end{array}
\right.
\end{array}\]
The triple $\three{\SF}{\fun}{\gr}$ is called the
{\em filter structure}
over $\calT$.  $\qed$
\end{enumerate}
\end{definition}

\noindent Notice that if $\{\tva\arr \tvb\mid\tvb\in f(\FFF\tva)\}$ is non-empty and
$\nu\in\con^\tytv$\footnote{By assumption $\nu\in\con^\tytv$ implies that
$\calT$ validates axiom $(\nu)$, see page~\pageref{ba}.} then $\nu\in
\FFF\{\tva\arr \tvb\mid\tvb\in f(\FFF\tva)\}$.

\noindent The definition of $\gr$, above,  appears natural once we
recall that axiom  ($\axtnu$) entails that $\nu$ is the universal type of functions.

Arrow types correspond to step functions, and they allow to describe the
functional behaviour of filters, in the following sense:

\begin{proposition}\label{fappp}
Let $\calT$ be a type theory.
For all $X\in\SF$ we get
$\fun(\FX) =
\DSup\{\step{\FFF\tva}{\FFF\tvb}\mid \tva\arr\tvb\in\FX\}$.
\end{proposition}

\begin{proof}
We show
$\tvd\in\fun(\FX)(\FFF\tvc)\ \Leftrightarrow\
\tvd\in(\DSup\{\step{\FFF\tva}{\FFF\tvb}\mid
\tva\arr\tvb\in\FX\})(\FFF\tvc).$\\
Let $\tvd\notteqt\tyt$, otherwise the thesis is trivial.\\[1em]
$\bar{ll}
\tvd\in X\cdot\FFF\tvc & \Leftrightarrow
\ear$\\
$\bar{ll}
\Leftrightarrow
&\exists I,\tva_i,\tvb_i. \tvc\tleqt\binto_{i\in I}\tva_i,
\binto_{i\in I} \tvb_i\tleqt\tvd\ \mbox{and}\
\forall i\in I. \tva_i\arr\tvb_i\in\FX\\
&\mbox{by definition of application and of $\tytv$-filter}\\
\Leftrightarrow &\exists I,\tva_i,\tvb_i. \step{\FFF\tvc}{\FFF\tvd}
\sqsubseteq\DSup_{i\in I}(\step{\FFF\tva_i}{\FFF\tvb_i}),
 \mbox{and}\ \forall i\in I. \tva_i\arr\tvb_i\in\FX\\
&\mbox{by definition of step function}\\
\Leftrightarrow & \step{\FFF \tvc}{\FFF \tvd}\sqsubseteq
\DSup\{\DSup_{i\in J}(\step{\FFF A_i}{\FFF B_i})\;\mid\;
A_i\to B_i\in X, J\ \mbox{finite set}\}\\
&\mbox{since $\step{\FFF\tvc}{\FFF\tvd}$ is compact
and the right-hand side is directed}\\
\Leftrightarrow& \step{\FFF\tvc}{\FFF\tvd}\sqsubseteq
\DSup\{\step{\FFF A}{\FFF B}\;\mid\; A\to B\in X\}\\
\Leftrightarrow &\tvd\in(\DSup\{\step{\FFF\tva}{\FFF\tvb}\mid
\tva\arr\tvb\in\FX\})(\FFF\tvc).
\ear$

\end{proof}

The next proposition provides a useful tool for relating arrow types
to application.

\begin{proposition} \label{simple} Let $\calT$ be a type
theory which validates $\bake$,
and let $\tyt\arr\tyt\in X$ if $\tyt\in\con^{\tytv}$,
then for all $X\in\SF, A,B\in \typetytv$ we get
\[\tvb\in X\cdot\FFF \tva \mbox{ iff } \tva\arr \tvb \in X.\]
\end{proposition}
\begin{proof}($\Rightarrow)$

If $\tvb\teqt\tyt$ then $\tyt\arr\tyt\tleqt
\tva\arr\tvb$ by rule ($\eta$).
So $\tva\arr \tvb \in X$ by assumption.
Otherwise,
by definition of application (Definition \ref{application}(\ref{application1})),
$\tvb\in X\cdot\FFF \tva$ iff
$\tvb\in\FFF\{\tvd\mid\exists\;\tvc\in\FFF\tva.\ \tvc\arr\tvd\in X\}$.

Then there is $I$ and $\tvc_i,\tvd_i$ such that
$\tva\tleqt\binto_{i\in I}\tvc_i$, $\binto_{i\in I}\tvd_i\tleqt\tvb$ and
$\tvc_i\arr\tvd_i\in X$ for all $i\in I$, by definition of $\tytv$-filter
(Definition \ref{filter}). So we get
$\tva\arr \tvb \in X$ by axiom ($\arr$-$\into$) and rule $(\eta)$. \\
($\Leftarrow)$ Trivial.
$\qed$
 \end{proof}

Filter structures induce immediately compositional $\lambda$-applicative structures.

\begin{proposition}\label{interpretation-in-filter-structures}
Let $\three{\SF}{\fun}{\gr}$ be a filter structure.

Let $\rho$ range  over the set of term environments ${\sf Env}_{\SF}=
[\mbox{\it Var} \rightarrow \SF]$.
Define the interpretation function: $\interpretation{\;}{\;}{\SF}:
\Lambda \times {\sf Env}_\SF \to \SF$   as follows:

\begin{itemize}
\item if there exists $x \in \mbox{\it Var}$ such that $\rho(x)=\emptyset$, then
$\interpretation{M}{\rho}{\SF}=\emptyset$;
\item otherwise put inductively:
$$\begin{array}{lll}

\interpretation{x}{\rho}{\SF} & = & \rho(x);\\
\interpretation{MN}{\rho}{\SF} & = &
\fun(\interpretation{M}{\rho}{\SF})(
\interpretation{N}{\rho}{\SF});\\
\interpretation{\lambda x.M}{\rho}{\SF} & = &
\gr(\lambda\!\!\lambda X\in\SF.\interpretation{M}{\rho[x:= X]}{\SF}).
\end{array}$$
\end{itemize}

The triple $\langle \SF, \cdot, \interpretation{\;}{\;}{\SF}\rangle $ is a compositional
$\lambda$-applicative structure.$\qed$
\end{proposition}

The interpretation function of a term coincides
with the set of types which are derivable for it.
This will be a crucial property in showing completeness using filter structures.

\begin{theorem}\label{ti-th}

For any $\lambda$-term
$M$ and environment $\en:\mbox{\it Var} \to \SF$,
\[ \interpretation{M}{\en}{\SF}=
\{\tva \in \typetytv \mid \exists \Gamma\ag\en.\;
\Gamma \ctas M:\tva\},\]
where $\Gamma\ag\en$ if and only if for all $x\in
\mbox{\it Var}$, $\rho(x)\neq \emptyset$, and moreover
$(x:\tvb)\in \Gamma$ implies  $\tvb\in\en(x).$
\end{theorem}

\begin{proof}
The thesis is trivial if $\rho(x)=\emptyset$ for some $x$. In such
a case
\[
\begin{array}{lllll}
\interpretation{M}{\en}{\SF} & = & \emptyset & = &
\{\tva \in \typetytv \mid \exists \Gamma\ag\en.\;
\Gamma \ctas M:\tva\},
\end{array}
\]
since no $\Gamma$ satisfies $\Gamma\ag\en$.

Otherwise we prove the thesis by induction on $M$.
Define\\
\indent $X_\nu=\mbox{ if }\nu\in\con^\tytv\mbox{ then }\FFF\nu \mbox{ else
} \emptyset$;\\
\indent
$X_\tyt=\mbox{ if }\tyt\in\con^\tytv\mbox{ then }\FFF\tyt \mbox{ else }
\emptyset$.\\
If $M\equiv x$, then\\
$\begin{array}{llll}
\interpretation{x}{\en}{\SF}& = &\en(x)&\\
& = & \{\tva\in \typetytv\mid \exists\tvb\in\en(x).\;\tvb\tleqt
\tva\}\\
& = & \{\tva\in \typetytv\mid \exists\tvb\in\en(x).\;x:\tvb\ctas
x:\tva\}&\mbox{by Theorem \ref{gen-l}(\ref{gen-l1}})\\
& = & \{\tva\in \typetytv\mid \exists\Gamma\ag\en.\;\Gamma\ctas
x:\tva\}.&\\
\end{array}$\\
If $M\equiv NL$, then\\
$\begin{array}{lll}
\interpretation{NL}{\en}{\SF}&=&\interpretation{N}{\en}{\SF}\cdot
\interpretation{L}{\en}{\SF}\\
& = &\FFF\{C\mid \exists B\in\interpretation{L}{\en}{\SF}.\;
B\to C\in\interpretation{N}{\en}{\SF}\}\\
&&\mbox{
by definition of application}\\
& = & \{\tva\in \typetytv\mid\exists I,
\tvb_i,\tvc_i.\;\tvb_i\arr\tvc_i
\in \interpretation{N}{\en}{\SF}, \tvb_i \in
\interpretation{L}{\en}{\SF},\\
&& \binto_{i \in I}\tvc_i\tleqt \tva\}\cup X_\tyt\\
& = & \{\tva\in \typetytv\mid\exists\Gamma\ag\en, I,
\tvb_i,\tvc_i.\;\Gamma\ctas
N:\tvb_i\arr\tvc_i,\\
&& \Gamma\ctas L:\tvb_i,
 \binto_{i \in I}\tvc_i\tleqt\tva\}\cup \{\tva\in
\type\mid A\teqt\tyt\}\\
&&\mbox{by induction, ($\mbox{\it weakening}$) and ($\lleq$)}\\
& = & \{\tva\in \typetytv\mid \exists\Gamma\ag\en.\; \Gamma\ctas NL:A\}\\
&&
\mbox{ by Theorem \ref{gen-l}(\ref{gen-l2}) and axiom $(\axtyt)$,
rule $(\tleqt)$}.
\end{array}$\\
If $M\equiv\lambda x.N$, then\\
$\begin{array}{lll}
\interpretation{\lambda x. N}{\en}{\SF}& = &
\gr(\lambda\!\!\lambda \FX\in\SF.\interpretation{N}{\en[x:=\FX]}{\SF})\\
& = & \FFF\{\tvb\arr\tvc\mid \tvc \in
\interpretation{N}{\en[x:=\FFF\tvb]}{\SF}\}\cup X_\nu\\
&&\mbox{by definition of }\gr\\
& = & \{\tva\in \typetytv\mid\exists\Gamma\ag\en, I,
\tvb_i,\tvc_i.\;
\Gamma, x:\tvb_i\ctas N:\tvc_i,\\
&& \binto_{i \in
I}\;(\tvb_i\arr\tvc_i)\tleqt \tva\}\cup \{A \in \typetytv\mid A\teqt\nu \}\\
&&\mbox{by induction, ($\mbox{\it weakening}$) and ($\lleq$)}\\
& = & \{\tva\in \typetytv\mid \exists\Gamma\ag\en.\;
\Gamma\ctas \lambda x.N:A\}\\
&&
\mbox{ by Theorem \ref{gen-l}(\ref{gen-l5}), axiom $(\axtnu)$, and
rule $(\tleqt).\;\;\qed$}
\end{array}$

\end{proof}

\section{Set-theoretic semantics of
 intersection-types}\label{completeness-theorem}

This is the main section of the paper. Here, we discuss completeness
for the three {\em set-theoretic semantics} of intersection-types
mentioned in the introduction.  In particular, we characterize those
intersection-type theories which induce complete type assignment
systems for the {\em inference}, the {\em simple} and the {\em F-}
semantics, over $\lambda$-applicative structures. As we will see these
conditions apply also to the preorders which induce complete systems with respect
to the three semantics, over $\lambda$-models.
We recall that according to these semantics the meaning of types
are {\em subsets} of the universe of discourse, i.e. the applicative structure.   The
``intersection'' type constructor is always interpreted as the {\em set-theoretic
intersection}. While, the ``arrow'' is interpreted as the set of those points, which
belong to a suitable distinguished set, and  whose applicative behavior is that of mapping
the antecedent of the arrow into the consequent.

As we remarked earlier the very existence of complete type assignment systems
for such semantics over applicative structures, is one of the
strongest motivations for the whole enterprise of developing a theory
of intersection-types.

In discussing completeness, {\em soundness} is not really an issue, since all
intersection-type theories are sound. To achieve {\em adequacy} and hence {\em
completeness} we have to restrict to two {\em disjoint} classes of type theories, namely
the  {\em natural\/} theories and the {\em strict} theories.
Filter structures
 are essential to showing adequacy. In such structures, in fact, the
 set-theoretic
interpretation of a type, as an appropriate subset,
is in one-to-one correspondence with the
{\em principal} filter generated by that type.

\begin{definition}\label{natural-type-theories}
\hfill
\begin{enumerate}
\item A type theory $\calT$ is {\em strict}  if
$\tyt \notin \con^\tytv$ and  it validates $\bake$ as defined in Figure \ref{f2}.
\item A type theory $\calT$ is {\em natural\/} if
$\tyt \in \con^\tytv$ and it validates $\abo$ as defined
in Figure \ref{f2}. $\qed$

\end{enumerate}
\end{definition}
Notice that by the blanket assumption, that axiom $(\nu)\in\tytv$ whenever
$\tnu\in\con^\tytv$ (cf. page~\pageref{ba}), a
strict type theory containing
the constant $\nu$ validates $\ehr$. All the theories appearing in Figure \ref{f2} are
natural, if they contain $\tyt$, and strict otherwise.

 Type interpretations can be given on $\lambda$-applicative structures
 once we have fixed a distinguished set of functional objects, $\Phi$. There are various
choices for this set. Amongst these there is a {\em maximal one} and a {\em minimal} one. The
former determines the {\em simple} semantics, the latter the {\em F-semantics}.

\begin{definition}[Type Interpretation Domain]\label{tyid}$_{}$
\begin{enumerate}
\item  \label{tyid12}A {\em type interpretation
domain\/}  is a qua\-dru\-ple
$\tid=\langle \D, \cdot, \interpretation{\;}{\;}{\D}, \Phi\rangle$ such that:
\begin{itemize}\item $\langle{\D},{\cdot}, \interpretation{\;}{\;}{\D}\rangle $ is  a
$\lambda$-applicative structure;
\item $\Phi$ is a subset of $\D$, called  the {\em functionality set}, such that
$ \interpretation{\lambda x. M}{\rho}{\D}\in \Phi$ for all $x,M,\rho$.
\end{itemize}
\item A type interpretation domain $\tid=\langle \D, \cdot,
\interpretation{\;}{\;}{\D}, \Phi\rangle$
is a {\em simple interpretation domain} if $\Phi=\D$;
\item A type interpretation domain $\tid=\langle \D, \cdot,
\interpretation{\;}{\;}{\D}, \Phi\rangle$
is an F-{\em interpretation domain} if $\Phi=\{ d \in \D
\mid d=\interpretation
{\lambda x. M}{\rho}{\D} \mbox{ for some }\ x,M,\rho \}$. $\qed$
\end{enumerate}
\end{definition}

\begin{definition}[Type Interpretation]\label{tyi}
Let $\tid=\langle \D, \cdot, \interpretation{\;}{\;}{\D},
\Phi\rangle$ be a  type interpretation domain.
 The {\em
type interpretation} induced by the type environment
$\ten:\con^{\tytv} \to {\sf P}(\D)$ is
defined by:
\begin{enumerate}
\item $\interpretation{\tyt}{\ten}{\tid}=\D$ ; \item
 $\interpretation{\nu}{\ten}{\tid}={\Phi}$;
\item $\interpretation{\tvav}{\ten}{\tid}=\ten(\tvav)$ if
$\tvav\in\con^{\tytv}$ and $\tvav\not=\tyt,\nu$;
\item $\interpretation{\tva\arr\tvb}{\ten}{\tid}=\{X\in
\Phi\;\mid\; \forall Y\in \interpretation{\tva}{\ten}{\tid}.
\; X\cdot Y \in\interpretation{\tvb}{\ten}{\tid}\}$;
\item $\interpretation{\tva\into\tvb}{\ten}{\tid}=\interpretation
{\tva}{\ten}{\tid}\cap\interpretation{\tvb}{\ten}{\tid}$. $\qed$
\end{enumerate}
\end{definition}

This definition is the counterpart for intersection-types of the {\em inference
semantics} for polymorphic types of \cite{mitc88}, generalized by allowing
$\langle {\D},{\cdot}, \interpretation{\;}{\;}{\D}\rangle $ to be just a $\lambda$-applicative
structure instead of a
$\l$-model.

Once we fix an applicative structure $\two{\D}{\cdot}$,
and an interpretation function
$\interpretation{\;}{\;}{\D}$, the above definition
depends on the choice of the functionality set $\Phi$
and the type environment
$\ten$.
The interpretation of the constants
$\{\tyt,\tnu\}$ takes into
account the corresponding axioms of the type assignment systems.\\

Notice that in the definition of $\l$-applicative
structure, we do not postulate, in general,
that $\interpretation{x}{\rho}{\D}=\rho(x)$.
Nevertheless the class of environments which
have this property will be of particular significance (provided
that they do not induce trivial interpretations, i.e. interpretations
in which all terms are equated).
Hence we put
\begin{definition}[Good environments] Let $\langle \D, \cdot,
\interpretation{\;}{\;}{\D}\rangle$ be a $\lambda$-applicative
structure.
 The term environment $\rho:
 \mbox{\it Var}\rightarrow \D$ is {\em good} if for all $x
\in \mbox{\it Var}$,
we have $\interpretation{x}{\rho}{\D}=\rho(x)$ and moreover there exist
two terms $M,N$ such that $\interpretation{M}{\rho}{\D}\neq
\interpretation{N}{\rho}{\D}$.
\end{definition}

In discussing {\em sound} type assignment systems we
consider only type interpretation
domains and type environments which are good (the notion
of goodness will depend on the current type theory and on
which kind of semantics we are considering)
and which agree with the inclusion relation
between
types in the following sense:

\begin{definition}\label{tyidg}
A type interpretation
domain ${\cal I}=\langle \D, \cdot, \interpretation{\;}{\;}{\D},
\Phi\rangle$
and a type environment $\ten:\con^{\tytv} \to {\sf P}(\D)$
\begin{enumerate}
\item\label{tyid-1} are $\tytv$-{\em good} if for all $\tva,\tvb \in
\type^\tytv$:
\begin{enumerate}
\item\label{tyid-1-1} for all {\em good} environments $ \rho$ and
$d\in \interpretation{\tva}{\cal V}{\cal I}$, $\rho[x:=d]$ is
{\em good};
\item\label{tyid-1-2}
for all {\em good} environments $ \rho$, terms $M$ and variables $x$,
  \[\forall d\in \interpretation{\tva}{\cal V}{\cal I}.\
  \interpretation{M}{\rho[x:=d]}{\D}\in
  \interpretation{\tvb}{\cal V}{\cal I}\;\;\Rightarrow\;
\interpretation{\lambda
x.M}{\rho}{\D}\in\interpretation{\tva\to\tvb}{\cal V}{\cal I};
\]
\end{enumerate}
\item\label{tyid-fg} are {\em $\tytv$-$F$-good\/} if they are $\tytv$-good
and moreover
for all {\em good} environments $ \rho$, variables $x$, and $A\in\type^\tytv$
\[\interpretation{x}{\rho}{\D}\in\interpretation{A}{\cal V}{\cal
I}\into \Phi\;\;\Rightarrow\;\;\interpretation{\l y.xy}{\rho}{\D}
\in \interpretation{A}{\cal V}{\cal I};\]
\item\label{tyid-2} {\em agree} with a type theory
$\calT$ iff for all
$\tva,\tvb\in\type^\tytv$:
$$ \tva\tleqt\tvb \ \Rightarrow \ \interpretation{\tva}{\cal V}{\cal
I}\subseteq
\interpretation{\tvb}{\cal V}{\cal I}.$$
\end{enumerate}
\end{definition}

Condition~(\ref{tyid-fg}) of Definition~\ref{tyidg} holds also when
$\cal I$ is an F-interpretation domain such that for all good $\rho$
we get that $\rho(x)\in\Phi$ implies  $\interpretation{x}{\rho}{\D}=
\interpretation{\l y.xy}{\rho}{\D}$.

Remark that the conditions~(\ref{tyid-1}) and~(\ref{tyid-fg}) of Definition~\ref{tyidg}
are true for all known models of (restricted) $\l$-calculus
(\cite{hindlong80}, \cite{plot75}, \cite{egidhonsronc92}, \cite{honsleni99}).

One can easily see that the following holds:

\begin{proposition}\label{agp}\hfill
\begin{enumerate}
\item \label{agp1}All type interpretation
domains and all type environments agree with $\abo$ and with $\ehr$.
\item  \label{agp2}All simple  interpretation
domains and  all type environments agree with $\bcd$. $\qed$
\end{enumerate}
\end{proposition}

We now introduce formally the three semantics. The definitions
follow in a natural way from
how we have argued informally so far,
but for the restriction
 (in the definition of $\agv $) to those type
 interpretation domains and type environments
which are $\tytv$-good ($\tytv$-F-good in the
 case of F-semantics) and which agree with $\calT$.

\begin{definition}[Semantic Satisfiability]
Let $\tid=\langle \D, \cdot, \interpretation{\;}{\;}{\D}, \Phi\rangle$
be
a  type interpretation domain.
\begin{enumerate}
\item
$\tid,\en,\ten \ag M:\tva$ iff $\interpretation{M}{\en}{\D}\in
\interpretation{\tva}{\ten}{\tid}$;
\item $\tid,\en,\ten \ag \Gamma$ iff $\tid,\en,\ten \ag x:\tvb$ for all
$(x\:\tvb)\in \Gamma$;
\item
$\Gamma \agv_i M:\tva$ iff $\tid,\en,\ten \ag \Gamma$ implies
$\tid,\en,\ten
\ag M:\tva$ for all $\tytv$-good type interpretation domains $\tid$ and
type environments $\ten$ which moreover agree with $\calT$, and
for all good term environments $\en$;
\item $\Gamma \agv_s M:\tva$ iff
$\tid,\en,\ten \ag \Gamma$ implies $\tid,\en,\ten
\ag M:\tva$ for all $\tytv$-good simple interpretations domains $\tid$ and
type environments $\ten$ which moreover agree with $\calT$, and
for all good term environments $\en$;
\item $\Gamma \agv_F M:\tva$ iff $\tid,\en,\ten \ag \Gamma$
implies $\tid,\en,\ten
\ag M:\tva$ for all $\tytv$-F-good F-interpretations domains $\tid$ and
type environments $\ten$ which moreover agree with $\calT$, and
for all good term environments $\en$.
\end{enumerate}
\end{definition}

For example $\not\agv_i x:\tyt\to\tyt$, $\agv_s x:\tyt\to\tyt$ and $\not\agv_F x:\tyt\to\tyt$.

In view of the above definition, we can say that the {\em inference semantics} is given by $\agv_i$, the {\em simple semantics} by $\agv_s$,
and the {\em F-semantics} by $\agv_F$. The following proposition is immediate:

\begin{proposition}
If $\Gamma \agv_i M:\tva$ then we have both $\Gamma \agv_s M:\tva$ and $\Gamma \agv_F
M:\tva$.
\end{proposition}

\begin{Notation} We shall denote with $\agv$ any of the three $\agv_i$, $\agv_s$, and
$\agv_F$.
\end{Notation}

Derivability
in the type system implies semantic satisfiability, as shown in the next theorem. Its proof
by induction on derivations is straightforward.

\begin{theorem}[Soundness]\label{soundness-of-type} $\Gamma \ctas
M:\tva$ implies $\Gamma \agv M:\tva$. $\qed$  \end{theorem}

\begin{proof} By induction on the derivation of $\Gamma \ctas
M:\tva$.\\
Rules ($\arE$) and $(\intI$) are sound by the definition of type
interpretation (Definition~\ref{tyi}).\\
The soundness of rule ($\arrI$) follows from the restriction to
$\tytv$-good type interpretation domains and type environments
(Definition~\ref{tyidg}(\ref{tyid-1})) and from the definition of
functionality set (Definition~\ref{tyid}(\ref{tyid12})).\\
Rule ($\tleqt$) is sound since we consider only type interpretation
domains and type environments which agree with $\sar{\tytv}$
(Definition~\ref{tyidg}(\ref{tyid-2})).
$\qed$\end{proof}

As regards to adequacy, first we observe that only  natural or strict  type theories can be
adequate.

\begin{proposition}(Adequacy implies naturality or strictness)\label{soundness-of-typead-nt}
If $\Gamma \agv
M:\tva$ implies $\Gamma \ctas M:\tva$ for all $\Gamma,M,A$,
then $\calT$ is a natural or a strict
type theory.
\end{proposition}
\begin{proof} It is easy to verify that
the hypothesis forces a type theory to validate
rule ($\eta$) and
 axiom ($\arr$-$\into$), and also axioms ($\tyt$), ($\tyt$-$lazy$)
when $\tyt\in\con^{\tytv}$, axiom ($\tnu$)
when $\tnu\in\con^{\tytv}$.
For instance, as regards to axiom ($\arr$-$\into$), consider the
$\tytv$-basis $\G = \set{x\:(A\to
B)\into (A\to C)}$. From Definition \ref{tyi} we get
$\G\agv x:A\arr B\into C$. Hence, by hypothesis, we have
$\G\ctas x:A\arr B\into C$. From Theorem
\ref{gen-l}(\ref{gen-l1}) it must hold
$(A\arr B)\into (A\arr C)\tleqt A\arr B\into C$, i.e.
axiom ($\arr$-$\into$) must hold.$\qed$
\end{proof}

Now we shall discuss adequacy for each of the three semantics separately.\\

First we consider the
inference semantics.
Our goal is to show that all natural and all strict  type theories are
adequate for the inference semantics.\\
For this proof, we focus on the applicative structure induced by the filter structure
$\three{\SF}{\fun}{\gr}$, and we put:

\begin{definition}\label{tid} Let $\calT$ be a  natural or strict type theory.
Let:
\begin{enumerate}
\item\label{tid1} $\Phi^\tytv$ be the functionality set defined by
\[
\Phi^\tytv =\left\{ \begin{array}{ll}
 \{\FX\in \SF\mid \tyt\to\tyt\in \FX\}& \mbox{if $\tyt\in\con^\tytv$;}\\
 \{\FX\in \SF\mid \tnu\in \FX\}& \mbox{if $\tnu\in\con^\tytv$;}\\
\SF&\mbox{otherwise.}
\end{array}
\right.
\]
\item\label{tid2} $\tid^\tytv$ be the type interpretation domain
$\four{\SF}{\cdot}{\interpretation{\;}{\;}{\tytv}}{\Phi^\tytv}$.
\item\label{tid3}  $\ten^{\tytv}:\con^{\tytv}\to{\sf P}(\SF)$ be
the type environment defined by
$$\ten^{\tytv}(\tvav)= \{\FX\in \SF\mid \tvav\in \FX\}.$$
\item $\interpretation{\;}{\;}{\tytv}:\typetytv\to{\sf P}(\SF)$ be the mapping
$\interpretation{\;}{\ten^{\tytv}}{\tid^{\tytv}}$. $\qed$
\end{enumerate}
\end{definition}
Notice that
\begin{itemize}
\item when $\tyt\in\con^{\tytv}$ we get  $\interpretation{\tyt}{\;}{\tytv}=
\SF= \{\FX\in \SF\mid \tyt\in \FX\}=\ten^\tytv(\tyt)$ and $\interpretation{\tyt\arr\tyt}{\;}{\tytv}=
\Phi^{\tytv}= \{\FX\in \SF\mid \tyt\to\tyt\in \FX\}$;
\item when $\tnu\in\con^{\tytv}$ we get $\interpretation{\tnu}{\;}{\tytv}=
\Phi^{\tytv}= \{\FX\in \SF\mid \tnu\in \FX\}=\ten^\tytv(\nu)$.
\end{itemize}
The mapping $\interpretation{\;}{\;}{\tytv}:\type^{\tytv}\to{\sf P}(\SF)$
has the property of associating to each type $\tva$
the set of filters which contain $\tva$
(thus preserving the property
through which we define
$\ten^{\tytv}$ in the basic
case of type constants).

\begin{lemma}\label{V0}
Let $\calT$ be a  natural or strict type theory then
$$\begin{array}{lll}
\ints{\tva}{\;} & = & \{\FX\in \SF\mid \tva\in \FX\}.
\end{array}$$

\end{lemma}
\begin{proof}
By induction on $\tva$. The only interesting case is  when $A$ is an arrow type. Remark that
if $X\in\SF$  but $X\notin \Phi^{\tytv}$ then all types in $X$ are intersections
of constant types. In fact if $X$ contains an arrow type, then it contains also
$\tyt\arr\tyt$ when $\tyt\in\con^{\tytv}$ (by rule ($\tyt$-$lazy$)), or
$\tnu$ when  $\tnu\in\con^{\tytv}$  (by rule ($\tnu$)),
so $X$ belongs to $\Phi^{\tytv}$.\\
If $\tva\equiv\tvb\arr\tvc$ we have
$$\begin{array}{llll}
\ints{\tvb\arr\tvc}{\;} & =
& \{\FX\in \Phi^\tytv\mid \forall \FY \in \ints{\tvb} {\;}\;\FX\cdot\FY
\in \ints{\tvc}{\;}\}&\mbox{by definition}\\
&=& \{\FX\in \Phi^\tytv\mid \forall \FY. \tvb\in\FY \Rightarrow
\tvc \in \FX\cdot\FY  \}&\mbox{by induction}\\
&=& \{\FX\in \Phi^\tytv\mid  \tvc \in \FX\cdot\FFF\tvb  \}&\mbox{by monotonicity}\\
&=& \{\FX\in \Phi^\tytv\mid  \tvb\arr\tvc \in  \FX  \}
&\mbox{by Proposition \ref{simple}}\\
&&&\mbox{and the definition of }\Phi^\tytv\\
&=& \{\FX\in \SF\mid  \tvb\arr\tvc \in  \FX  \}
&\mbox{by above.$\qed$}
\end{array}$$
\end{proof}

\newcommand{\intsu}[1]{\dlsqb{#1}\drsqb^{\tytv}}
\begin{lemma}\label{goodl}
Let $\calT$ be a natural or strict  type theory. Then
$\tid^\tytv$, $\ten^{\tytv}$ are $\tytv$-good and they agree
with $\calT$.
\end{lemma}
\begin{proof}
$\tid^\tytv$, $\ten^{\tytv}$ satisfy condition (\ref{tyid-1-1})
of Definition~\ref{tyidg} since $\emptyset\notin\intsu{A}$
for all $A\in\type^\tytv$ by Lemma~\ref{V0}.\\
For condition~(\ref{tyid-1-2}) of Definition~\ref{tyidg} let
$X\in \intsu{A}$ be such that
$\interpretation{M}{\rho[x:=X]}{\tytv}\in\intsu{B}$.
Then, by Lemma \ref{V0}, $B\in\interpretation{M}{\rho[x:=X]}{\tytv}$,
hence $B\in f(X)$, where we have put
$f=\ll d.\interpretation{M}{\rho[x:=d]}{\tytv}$.
Notice that:

$\bar{llll}
\fun(\gr(f))& = &\DSup\{\step{\FFF\tva}{\FFF\tvb}\mid
\tva\arr\tvb\in\gr(f)\}&
  \mbox{\ by Proposition \ref{fappp}}\\
& \Dgeq & \DSup\{\step{\FFF\tva}{\FFF\tvb}\mid \tvb\in f(\FFF\tva)\}&\mbox{\ by definition of $\gr$} \\
& = &f&\mbox{\ by definition of step function.}\\
\ear$

\noindent
We are done since $\fun(\gr(f))(X) =
\interpretation{\l x.M}{\rho}{\tytv}\cdot X$.

Lastly notice that as an immediate consequence of the Lemma~\ref{V0} we get
$$A\tleqt B \Leftrightarrow \forall X\in \SF.[ A\in X\Rightarrow B\in X
]
\Leftrightarrow \ints{A}\;\subseteq \ints{B},$$
and therefore $\tid^\tytv, \ten^\tytv$ agree with $\sar{\tytv}$.$\qed$
\end{proof}

Finally we can prove the desired adequacy result.

\begin{theorem}(Naturality or strictness imply adequacy) \label{nia}
Let $\calT$ be a natural or a strict type theory. Then
$\Gamma \agv_i M:\tva$  implies  $\Gamma \ctas M:\tva$.
\end{theorem}
\begin{proof} We consider the type interpretation domain
${\cal I}^\tytv$.
Let $\en_{\Gamma}$ be the term environment
defined by $\en_{\Gamma}(x)=
\{\tva \in \typetytv \mid
\Gamma \ctas x:\tva\}$. It is easy to verify that $\tid^\tytv,
\en_{\Gamma},  \ten^\tytv\ag \Gamma$ and that
for all $\Gamma'\ag \en_{\Gamma}$ we have  $\Gamma'\ctas M:A
\Rightarrow\Gamma\ctas M:A$. Hence we have:

$$\begin{array}{llll}
\Gamma \ag^\tytv_i M:\tva&\Rightarrow&
\interpretation{M}{\en_{\Gamma}}{\tytv}\in
\ints{\tva}{\;}&
\mbox{by Lemma~\ref{goodl} since }\tid^\tytv, \en_{\Gamma},
\ten^\tytv\ag \Gamma\\
&\Rightarrow&\tva\in
\interpretation{M}{\en_{\Gamma}}{\tytv}&\mbox{by Lemma \ref{V0}}\\
&\Rightarrow&\Gamma \ctas M:\tva&\mbox{by Theorem \ref{ti-th} and}\\
&&&\mbox{the
above property.} \
\qed
\end{array}$$
\end{proof}

Hence the only type theories which turn out to be complete with respect to the inference
semantics are the natural and the strict type theories. There are of course many
theories of interest which do not belong to these classes. For instance the intersection-type
theory which induces the filter structure isomorphic to Scott's ${\cal P}(\omega)$ is
such a theory. The reader can see \cite{bare00} for more examples.

Notice that the theories $\Sigma^{\abo}$, $\Sigma^{\bcd}$ induce filter structures which
are $\lambda$-models \cite{barecoppdeza83}, the theory
$\Sigma^{\ehr}$ induces a model for the  $\lambda_v$-calculus \cite{egidhonsronc92}, and the theory
$\Sigma^{\bake}$ induces a model for the  $\lambda$-$\I$-$\N$-calculus \cite{honsleni99}. Hence  we have
that natural theories, which induce $\lambda$-models,  are complete also for the class of $\lambda$-models, and strict
theories, which induce models of the other two restricted
$\lambda$-calculi, are complete for the  corresponding classes of models.

Now we characterize those theories which are complete
with respect to  the simple semantics.

\begin{theorem}(Adequacy for the simple semantics) \label{st}
$\Gamma \agv_s M:\tva$  implies  $\Gamma \ctas M:\tva$
iff $\calT$ is a strict type theory such that
$\tnu\notin\conv$
or a natural type theory which validates axiom
($\tyt$-$\eta$).
\end{theorem}
\begin{proof}($\Rightarrow$) From  Proposition
\ref{soundness-of-typead-nt} it follows that $\calT$ is
natural or strict.\\
If $\tnu\in\conv$ we have, for any type interpretation
domain ${\cal I} =
\four{\D}{\cdot}{\interpretation{\;}{\;}{\D}}{\Phi}$ and
$\cal V$ type environment:
$\interpretation{\tnu}{\cal V}{\cal I} = \Phi = \D$, hence
$\agv_s x:\tnu$. But it never holds $\vdash^\tytv_\nu x:\tnu$ by
Theorem~\ref{gen-l}(\ref{gen-l1}),
hence simple adequacy fails if $\tnu\in\conv$.\\
It is easy to check that if
$\tyt\notteqt\tyt\to\tyt$ then simple adequacy fails for
$\tityt^{\tytv}$. We have $ \agv_s x:\tyt\to\tyt$
since $\interpretation{x}{\rho}{\D}\cdot d\in\D$
for all $\D,d\in\D$ and $\rho:\env_\D\to\D$.
By
Theorem~\ref{gen-l}(\ref{gen-l1}) we can deduce  $\ctas_\tyt x:\tyt\to\tyt$ only if
$\tyt\teqt\tyt\to\tyt$.\\
This proves ($\Rightarrow$).

\noindent ($\Leftarrow$).
To prove that $\Gamma \agv_s M:\tva$  implies
$\Gamma \ctas M:\tva$ under the given conditions we use
the simple type interpretation domain
$\four{\SF}{\cdot}{\interpretation{\;}{\;}{\tytv}}{\SF}$, which
is just $\tid^\tytv$ as defined in Definition
\ref{tid},
with
$\Phi^\tytv = \SF$, being either $\nu\notin\conv$ or $\tyt\sim_\tytv
\tyt\arr\tyt$. By Lemma \ref{V0} it follows
$\ints{A}{} = \{\FX\in \SF\mid \tva\in \FX\}$. So we have that
$\G\agv_s M:A$ implies $A\in\interpretation{M}{\rho_\G}{\tytv}$
and we conclude $\G\ctas M:A$ as in the last step of the
proof of Theorem \ref{nia}.
$\qed$ \end{proof}

Among the type theories of Figure~\ref{f2}, those adequate for the
simple semantics are $\Sigma^{\bake}$ and $\Sigma^{\bcd}$. Other adequate type theories
in the literature are those of
\cite{honsleni99,egidhonsronc92,scot72,park76,coppdezazacc87,honsronc92}. Non
adequate type theories are all those inducing computationally adequate
models for the lazy $\lambda$-calculus, e.g. $\Sigma^{\abo}$, or
for the call-by-value $\lambda$-calculus, e.g. $\Sigma^{\ehr}$. The same argument
used for the inference semantics allows to show that the natural
theories mentioned in Theorem \ref{st}, which induce $\lambda$-models,  are precisely those which are
complete also for the class of $\lambda$-models, and the
strict theories, which induce  $\lambda_v$-models and
$\lambda$-$\I$-$\N$-models,  are complete for the corresponding classes of models.  The completeness for the simple
semantics of $\ti_\tyt^\tytv$ whenever $\sar{\tytv}$ validates $\bcd$ and
$\three{\SF}{\cdot} {\interpretation{\;}{\tytv}{\;}}$ is a $\l$-model
was proved in \cite{coppdezahonslong84} using filter models and in
\cite{coppdezazacc87} using the term model of $\beta$-equality.  \\

Finally we turn to the F-semantics.
The following definition singles out the type theories which are adequate
for the F-semantics as proved
in Theorem~\ref{Ft}.

\begin{definition}\label{fsem}
A type theory $\calT$ is an F-{\em type theory} iff
\begin{enumerate}
\item \label{fsem1} either $\calT$ is a strict or a natural type theory
such that $\nu\notin\con^\tytv$ and for all $\psi\in\con^\tytv$,
$\tva,\tvb\in\type^\tytv$, there are
$I, A_i,B_i\in\type^\tytv$ such that
\indent $\psi\into(A\to B)\teqt\binto_{i\in I}(A_i\to B_i)$;
\item \label{fsem2} or $\calT$ is a strict type theory such that
$\nu\in\con^\tytv$ and for all $\psi\in\con^\tytv$
either $\tnu\tleqt\psi$ or there are
$I, A_i,B_i\in\type^\tytv$ such that
\indent $\psi\into\nu\teqt\binto_{i\in I}(A_i\to B_i)$.
 $\qed$
\end{enumerate}
\end{definition}
For example $ \sar{\bake}$, $ \sar{\ehr}$, and $\sar{\abo}$ are
F-type theories.\\
Notice that a natural type theory $\sar{\tytv}$ which validates axiom
$(\tyt$-$\eta)$ is an F-type theory iff for all $\psi\in\con^\tytv$ we get
$\psi\teqt\binto_{i\in I}(A_i\to B_i)$, for some
$I, A_i,B_i\in\type^\tytv$.\\

Next lemma shows that all types of a F-type theory satisfy the conditions of previous
definition.

\begin{lemma}\label{f-type-theory-simple-lemma}
Let $\sar{\tytv}$ be a F-type theory. Then
\begin{enumerate}
\item\label{f-type-1}
if $\tnu\notin\conv$, then
for all $\tva,\tvb,C\in\type^\tytv$,
$C\into(A\to B)\teqt\binto_{i\in I}(A_i\to B_i)$,
for some $I, A_i,B_i\in\type^\tytv$;
\item \label{f-type-2} if
$\nu\in\con^\tytv$, then for all $\tvc\in\type^\tytv$
either $\tnu\tleqt\tvc$ or
$\tvc\into\nu\teqt\binto_{i\in I}(A_i\to B_i)$, for some
$I, A_i,B_i\in\type^\tytv$;
\item \label{f-type-3} for all $\tva,\tvb,\tvc\in\type^\tytv$,
$\tvc\into(A\to B)\teqt\binto_{i\in I}(A_i\to B_i)$, for some
$I$, $A_i,B_i\in\type^\tytv$.
\end{enumerate}
\end{lemma}

\begin{proof}
We just prove the more difficult case, namely (\ref{f-type-2}).
We reason by induction on the structure of $C$. If
$C\in\con^\tytv$ the thesis is trivial. If $C\equiv D\arr E$, then
$C\tleqt \tnu$, hence $C\into \tnu\teqt D\arr E$. If
$C \equiv D\into E$ and $\tnu\tleqt C$ then
the thesis is immediate. Otherwise we cannot have both $\tnu\tleqt D$
and $\tnu\tleqt E$. Let us suppose $\tnu\not\tleqt D$. Then, by induction,
it follows $D\into \tnu\teqt \binto_{i\in I}(A_i\to B_i)$,
for suitable $I$ and $A_i, B_i\in\type^\tytv$. Now,
if $\tnu\tleqt E$, we get $C\into \tnu\teqt D\into \tnu$ and
we are done by above. If $\tnu\not\tleqt E$, then, by induction,
it follows $E\into \tnu\teqt \binto_{j\in J}(A'_j\to B'_j)$ for
suitable $J$ and $A'_j, B'_j\in\type^\tytv$. Therefore
$C\into \tnu\teqt(\binto_{i\in I}(A_i\to B_i))\into(\binto_{j\in J}(A'_j\to
B'_j))$.$\qed$
\end{proof}

To discuss F-semantics it is useful to characterize the subset of types which are functional.

\begin{definition}\label{ffun}
We define the predicate $\ffun$ on $\type^\tytv$ by induction on the structure of types:
\begin{enumerate}
\item $\ffun(\psi)=\psi\teqt\nu\;{\sf or}
\;\psi\teqt\binto_{i\in I}(A_i\to B_i)$ for some
$I, A_i,B_i\in\type^\tytv$; \item $\ffun(A\to B)={\sf true}$;
\item $\ffun(A\into B)=\ffun(A)\;{\sf or}\; \ffun(B)$.
$\qed$
\end{enumerate}
\end{definition}

The following proposition gives an alternative characterization
of functional types for F-type theories.
\begin{proposition}\label{ffunp}
If $\calT$ is an F-type theory then $\ffun(A)$ iff either $A\teqt\nu$ or
$A\teqt\binto_{i\in I}(A_i\to B_i)$ for some $I,
A_i,B_i\in\type^\tytv$.$\qed$
\end{proposition}

\begin{proof}
($\Leftarrow$) is trivial.

($\Rightarrow$) We reason by induction on the structure of $A$.
If $A\in\con^\tytv$, or $A\equiv B\to C$, or
$A\teqt \tnu$, the thesis follows by
definition of $\ffun(A)$. Otherwise we have $A\equiv B\into C$
and either $\ffun(B)$ or $\ffun(C)$. We
assume $\ffun(B)$, the case $\ffun(C)$ being similar.
By induction either
$B\teqt\tnu$ or $B\teqt\binto_{i\in I}(A_i\to B_i)$ for
some $I$ and $A_i, B_i\in\type^\tytv$.
In the first case either $\nu\tleqt C$ and $A\teqt C\into\nu \teqt\nu$ or
$A\teqt C\into\nu \teqt
\binto_{i\in J}(A'_j\to B'_j)$, for some $J$ and $A'_j, B'_j\in\type^\tytv$ by
Lemma~\ref{f-type-theory-simple-lemma}(\ref{f-type-2}).
In the second case $A\teqt C \into \binto_{i\in I}(A_i\to B_i)$. By choosing an
arbitrary $i\in I$ we get $C \into (A_i\to B_i)\teqt \binto_{i\in J}(A'_j\to B'_j)$, for
some $J$ and $A'_j, B'_j\in\type^\tytv$ by
Lemma~\ref{f-type-theory-simple-lemma}(\ref{f-type-3}). So we conclude
$A\teqt (\binto_{i\in I}(A_i\to B_i))\into(\binto_{i\in J}(A'_j\to B'_j))$.

\end{proof}

To prove adequacy we will again use the filter structure
$\three{\SF}{\fun}{\gr}$ for defining, as in the
previous cases, the type interpretation
domain $\four{\SF}{\cdot}{\interpretation{\;}{\;}{\tytv}}{\Phi^\tytv_F}$.
The definition below differs from Definition \ref{tid}
in that we choose a different functionality set.


\begin{definition}\label{fsemd}
Let $\calT$ be an F-type theory. Let:
\begin{enumerate}
\item $\Phi^\tytv_F$ be the functionality set defined by
\[
\Phi^\tytv_F =\{ X \in \SF \mid X=\interpretation
{\lambda x. M}{\rho}{\tytv} \mbox{ for some }\ x,M,\rho \};
\]
\item $\tid^\tytv_F$ be the type interpretation domain
$\four{\SF}{\cdot}{\interpretation{\;}{\;}{\tytv}}{\Phi^\tytv_F}$;
\item  $\ten^{\tytv}_F:\con^{\tytv}\to{\sf P}(\SF)$ be
the type environment defined by
$$\ten^{\tytv}_F(\tvav)=
\{\FX\in \SF\mid \tvav\in \FX\};$$
\item $\interpretation{\;}{F}{\tytv}:\typetytv\to{\sf P}(\SF)$ be the mapping
$\interpretation{\;}{\ten^{\tytv}_F}{\tid^{\tytv}_F}$. $\qed$
\end{enumerate}
\end{definition}

When restricting to F-type theories, all filters which contain a
functional type belong to
the functionality set.

\begin{lemma}\label{ffunl}
Let $\calT$ be an F-type theory and $X\in\SF$. Then $A\in X$
and $\ffun(A)$ imply $X\in\Phi^\tytv_F$.
\end{lemma}
\begin{proof}
We show that under the given conditions
$$X=\interpretation{\l y.xy}{\rho_0}{\tytv}\mbox{ where }
\rho_0(x)=X.$$
Proof of $X\subseteq\interpretation{\l y.xy}{\rho_0}{\tytv}$.
Take an arbitrary $B\in X$.
Notice that if $\nu\in\con^\tytv$ then $\ffun(A)$ implies $A\tleqt\nu$
by Proposition~\ref{ffunp}.
Moreover $\ffun(A)$ implies $\ffun(A\into B)$ by Definition~\ref{ffun}.
Then either
$A\into B\teqt\nu$ or
$A\into B\teqt\binto_{i\in I}(C_i\to D_i)$ for some
$I, C_i,D_i\in\type^\tytv$
again by Proposition~\ref{ffunp}. In the first case we get
$\nu\tleqt B$ and then
$\ctastnu \l y.xy:B$ by axiom $(\axtnu)$ and rule $(\tleqt)$.
In the second case we can derive
$\set{x\:\binto_{i\in I}(C_i\to D_i)}\ctas \l y.xy:\binto_{i\in I}(C_i\to
D_i)$
using axiom $(\ax)$ and rules $(\tleqt)$, $(\arE)$, $(\arrI )$,
and $(\intI)$. This implies
$\set{x\:A\into B}\ctas \l y.xy:B$ by rules $(\tleqt)$ and $(\lleq)$.
In both cases we conclude
$B\in\interpretation{\l y.xy}{\rho_0}{\tytv}$
by Theorem \ref{ti-th},
since $\emptyset\ag\rho_0$ (case $A\into B\teqt\nu$) and
$\{x\:A\into B\} \ag \rho_0$ (case
$A\into B\teqt\binto_{i\in I}(C_i\to D_i)$).\\
Proof of $\interpretation{\l y.xy}{\rho_0}{\tytv}\subseteq X$.
By Theorem~\ref{ti-th}, $B\in\interpretation{\l y.xy}{\rho_0}{\tytv}$
implies $\set{x\:C}\ctas  \l y.xy:B$ for some
$C\in X$. If $\tyt\in\con^\tytv$ and
$B\teqt \tyt$ then $B\in X$ for all $X$. If $\nu\in\con^\tytv$ and
$\nu\tleqt B$ then $B\in X$ since $A\tleqt\nu$ by
Proposition~\ref{ffunp}.
Otherwise
we get  $\set{x\:C,y\:D_i}\ctas  xy:E_i$
for some $I, D_i,E_i\in\type^\tytv$ such that
$\binto_{i\in I}(D_i\to E_i)\tleqt B$ by
Theorem~\ref{gen-l}(\ref{gen-l5}).
This implies $\set{x\:C,y\:D_i}\ctas  x:F_i\to E_i$,
$\set{x\:C,y\:D_i}\ctas  y:F_i$ for some
$F_i\in\type^\tytv$ by Theorem~\ref{gen-l}(\ref{gen-l4}).
Using Theorem~\ref{gen-l}(\ref{gen-l1})
we have $C\tleqt F_i\to E_i$ and $D_i\tleqt F_i$ for all $i\in I$,
so we get $C\tleqt D_i\to E_i$ by rule $(\eta)$, and we can conclude
$C\tleqt B$, i.e. $B\in X$.
$\qed$
\end{proof}

\begin{lemma}\label{V1}
Let $\calT$ be a F-type theory then
$$\begin{array}{lll}
\interpretation{A}{F}{\tytv} & = & \{\FX\in \SF\mid \tva\in \FX\}.
\end{array}$$
\end{lemma}
\begin{proof}
The proof by induction on $A$ is similar to that of Lemma \ref{V0}. All cases are trivial
but  for $\tnu$ and arrow types.\\ If $A\equiv\tnu$
let $X$ be any filter in $\Phi_F^\tytv$, that is $X=\interpretation{\l
x.M}{\rho}{\tytv}$ for some $x,M,\rho$. Then, by Theorem \ref{ti-th},
$X=\{B\in\type^\tytv\mid\exists\G\ag\rho.\;\G\ctastnu \l x.M:B\}$.
Since $\ctastnu \l x.M:\tnu$, we have $\tnu\in X$.
Vice versa, if $\tnu\in X$, then by Definition~\ref{ffun} $\ffun(\nu)$, and so by
Lemma~\ref{ffunl},
$X\in\Phi_F^\tytv$. We have proved, when $\tnu\in\conv$, that
\[X\in \Phi_F^\tytv\;\Leftrightarrow \tnu\in X.\]
Hence $\interpretation{\tnu}{F}{\tytv} = \Phi_F^\tytv =
\{X\in\SF\mid \tnu\in X\}$.\\
If $A\equiv B\to C$ we have
$$\begin{array}{llll}
\interpretation{\tvb\arr\tvc}{F}{\tytv} & =
& \{\FX\in \Phi_F^\tytv\mid  \tvc \in \FX\cdot\FFF\tvb  \}&\mbox{as
in the proof of Lemma \ref{V0}}\\
&=& \{\FX\in \Phi_F^\tytv\mid  \tvb\arr\tvc \in  \FX  \}
&\mbox{by Proposition \ref{simple} since,}\\
&&&\mbox{when $\tyt\in\con^{\tytv}$, Theorem \ref{ti-th}}\\
&&&\mbox{and $X=\interpretation{\l x.M}{\rho}{\tytv}$ imply
$\tyt\to\tyt\in X$}\\
&=& \{\FX\in \SF\mid  \tvb\arr\tvc \in  \FX  \}
&\mbox{by Lemma \ref{ffunl} since }\ffun(B\to C) .\qed
\end{array}$$
\end{proof}

\begin{lemma}\label{good2}
Let $\calT$ be an F-type theory. Then
$\tid_F^\tytv$, $\ten_F^{\tytv}$ are $\tytv$-F-good and they agree
with $\calT$.
\end{lemma}
\begin{proof}
We can mimick the proof of Lemma \ref{goodl}, using
Lemma \ref{V1} instead of Lemma \ref{V0}, for
all points of Definition \ref{tyidg} but for
(\ref{tyid-fg}). So we are left to prove that this last condition
holds. The key observation is that by Lemma~\ref{V1} and Theorem~\ref{ti-th}
$$(*)\;\;\interpretation{M}{\rho}{\tytv}\in\interpretation{A}{F}{\tytv}\iff
\G\vdash^\tytv M:A\mbox{ for some $\tytv$-basis $\G$ such that $\G\ag\rho$.}$$
Let $\interpretation{x}{\rho}{\tytv} \in
\interpretation{A}{F}{\tytv}\into \Phi_F^\tytv$.
Then
$\interpretation{x}{\rho}{\tytv} = \interpretation{
\l z.M}{\rho'}{^\tytv}$ for some $ z,M,\rho'$. By $(*)$
there exists a $\tytv$-basis $\G'$ such that $\G'\ag\rho'$ and
$\G'\vdash^\tytv \l z.M:A$. By Theorem~\ref{gen-l}(\ref{gen-l5}) there exist
$I, A_i, B_i$, such that $\G'\vdash \l z.M:\binto_{i\in I}
(A_i\to B_i)$ and $\binto_{i\in I}
(A_i\to B_i)\tleqt A$. Hence by $(*)$
$\interpretation{\l z.M}{\rho'}{\tytv}
\in \interpretation{\binto_{i\in I}(A_i\to B_i)}{F}{\tytv}$. Since
$\interpretation{x}{\rho}{\tytv} = \interpretation{
\l z.M}{\rho'}{^\tytv}$ by $(*)$
there exists a $\tytv$-basis $\G\ag\rho$, such that
$\G\vdash^\tytv x:\binto_{i\in I}(A_i\to B_i)$, hence
$\G\vdash^\tytv \l y.xy: \binto_{i\in I}(A_i\to B_i)$,
by applying rules ($\tleqt$), ($\to$E), ($\to$I) and ($\into$I).
So we have by $(*)$ $\interpretation{\l y.xy}{\rho}{\tytv}
\in \interpretation{\binto_{i\in I}(A_i\to B_i)}{F}{\tytv}$.
Since ${\cal I}^\tytv_F$, ${\cal V}^\tytv_F$
agree with $\sar{\tytv}$, we get $\interpretation{\binto_{i\in I}(A_i\to
B_i)}{F}{\tytv}\subseteq
\interpretation{A}{F}{\tytv}$, so we conclude
$\interpretation{\l y.xy}{\rho}{\tytv}\in
\interpretation{A}{F}{\tytv}$.$\qed$
\end{proof}

\begin{theorem}(Adequacy for the F-semantics) \label{Ft}
$\Gamma \agv_F M:\tva$  implies  $\Gamma \ctas M:\tva$
iff $\calT$ is an F-type theory.
\end{theorem}
\begin{proof}
First we check that the given conditions are necessary.\\ Let
${\interpretation{x}{\rho}{\D}}\in
{\interpretation{\psi\into(A\to B)}{\ten}{\D}}$ for some $\tid=\langle
\D, \cdot, \interpretation{\;}{}{\D}, \Phi\rangle$,
$\ten$ which are F-$\tytv$-good, agree with $\sar{\tytv}$, and some
good $\rho$. Then
${\interpretation{x}{\rho}{\D}}\in\Phi$, since $\interpretation{A\to B}{\cal
V}{\D}\subseteq \Phi$. By Definition \ref{tyidg}(\ref{tyid-fg}) it follows
$\interpretation{\l y.xy}{\rho}{\D}\in\interpretation{\psi\into
(A\to B)}{\cal V}{\D}$, hence
$x\:\psi\into(A\to B)\agv_F
\lambda y.xy:\psi\into(A\to B)$.
By a similar argument we can obtain $x\:\psi\into\nu\agv_F
\lambda y.xy:\psi\into\nu$
when $\nu\in\con^\tytv$. Therefore we have
F-adequacy of ${\ti}^{\tytv}$ only if we can prove
$x:\psi\into(A\to B)\ctas \lambda y.xy:\psi\into(A\to B)$
(respectively
$x:\psi\into\nu\ctas_\nu \lambda y.xy:\psi\into\nu$
when $\tnu\in\con^\tytv$). Let $\tnu\notin\con^\tytv$.
\\[1em]
$\begin{array}{ll}
x\:\psi\into(A\to B)\ctas \lambda y.xy:\psi\into(A\to B)&
\Rightarrow
\end{array}$\\
$\begin{array}{ll}
\Rightarrow & x\:\psi\into(A\to B), y\:A_i\ctas  xy:B_i\\
&\mbox{for some $I, A_i,B_i\in\type^\tytv$ such that
$\binto_{i\in I}(A_i\to B_i)\tleqt\psi\into(A\to B)\;(\dagger)$,}\\
&\mbox{by Theorem~\ref{gen-l}(\ref{gen-l5})} \\
\Rightarrow&\mbox{$x\:\psi\into(A\to B), y\:A_i\ctas  x:C_i\to
B_{i}$ and $x\:\psi\into(A\to B), y\:A_i\ctas  y:C_i$}\\
&\mbox{for some $ C_i\in\type^\tytv$, by Theorem
\ref{gen-l}(\ref{gen-l4})}\\
\Rightarrow&\mbox{$\psi\into(A\to B)\tleqt\binto_{i\in I}(C_{i}\to
B_{i})$
and $\tva_i\tleqt C_i$}\\
&\mbox{by Theorem~\ref{gen-l}(\ref{gen-l1})}\\
\Rightarrow&
\mbox{$\psi\into(A\to B)\tleqt \binto_{i\in I}(A_i\to B_i)$}\\
&\mbox{by rule ($\eta$).}\\
\end{array}$

\noindent
This last judgment along with $(\dagger)$ implies
$\psi\into(A\to B)\teqt \binto_{i\in I}(A_i\to B_i)$. \\
Similarly from $x\:\psi\into\nu\agv_F \lambda y.xy:\psi\into\nu$ we can
show
that either $\tnu\tleqt\psi$ or
$\psi\into\nu\teqt\binto_{i\in I}(A_i\to B_i)$ when
$\nu\in\con^\tytv$.\\
For the vice versa, we consider the F-interpretation domain
${\cal I}_F^\tytv$ and the type environment ${\cal V}_F^\tytv$
of Definition~\ref{fsemd}.
They are $\tytv$-F-good and agree with $\calT$ by Lemma
\ref{good2}. By Lemma \ref{V1}
$\interpretation{A}{F}{\tytv}=\{X\in\SF\mid A\in X\}$.
So we have that $\G\agv_F M:A$ implies
$\G\ctas M:A$ mimicking the proof of Theorem~\ref{nia}.$\qed$
\end{proof}

The theories $\Sigma^{\bake}$, $\Sigma^{\ehr}$, and $\Sigma^{\abo}$, as well as
the type theories of
\cite{honsleni99,scot72,park76,coppdezazacc87,honsronc92} are adequate for the F-semantics.
Moreover for the last five
the simple semantics coincides with the F-semantics. The theory $\Sigma^{\bcd}$ is an example of  a
theory which is not adequate with respect to the F-semantics.
The remark concerning $\lambda$-models and restricted $\lambda$-models made for the
inference and the simple semantics, applies also to the F-semantics.

\section{Related work and final remarks}\label{fin}

In the literature there are essentially five ways of interpreting Curry's
types in a model of the untyped $\lambda$-calculus. They differ in the
interpretation of the {\em arrow} type constructor.
In what follows we shall mainly  follow the terminology of~\cite{hind83}.

The simple and the F-semantics are defined as expected.

Following~\cite{scot80}, the {\em quotient set semantics} takes into account that
we want to consider equivalent two functions iff they give equivalent results
when applied to equivalent arguments. So types are interpreted as partial equivalence
relations  of the domain rather than simply as subsets. The arrow constructor is defined as
for logical relations:
$d\sim_{A\to B}d'$ iff for all $c,c'$ such that  $c\sim_{A}c'$ it holds
$d\cdot c\sim_{B}d'\cdot c'$.

The {\em F-quotient set semantics}~\cite{scot76}, modifies  the quotient set
semantics, in the same way as the F-semantics modifies the simple semantics. Namely it
requires that all elements of the domain which are equivalent with respect to an arrow must
be canonical representatives  of functions.

Finally, Mitchell in~\cite{mitc88} introduces another semantics, which he calls
{\em inference semantics}, in which the interpretation of the arrow
must {\em at least} contain the canonical representatives
of functions which behave correctly with respect to
the application.

All the above semantics easily extend to {\em intersection-types}
~\cite{barecoppdeza83},
\cite{dezamarg86} and to {\em polymorphic types}~\cite{mitc88}.

The crucial question in the semantics of types is the completeness of
type assignment systems. Hindley proved in~\cite{hind83} that Curry's
type assignment system is complete for all the mentioned semantics.
More specifically \cite{hind83} and~\cite{hind83a} show the completeness for the simple
semantics and moreover that:
\begin{enumerate}
\item \label{a1} $\Gamma\agv_F M:\tva$ if and only if $\Gamma\agv_s M:\tva$, when
 $\tva$ is a Curry type;
\item \label{a2} the simple semantics is a particular case of
the quotient set semantics;
\item \label{a3} the F-semantics is a particular case of the F-quotient set semantics.
\end{enumerate}

The argument showing points (\ref{a2}) and (\ref{a3}) easily extends to
intersection and polymorphic types, so for these type disciplines it is enough to discuss
only completeness for the simple and the F-semantics to get completeness results for the
quotiented versions.
One could define  also a quotient version of the inference semantics, but this would be
treated similarly.

The completeness
with respect to the simple semantics,
of various intersection-type assignment systems, over  $\lambda$-models,  has been proved
in
\cite{barecoppdeza83,hind82,coppdezahonslong84,coppdezazacc87,bake92}.

As far as the completeness with respect to the F-semantics
of intersection-type assignment systems over  $\lambda$-models, we can
cite~\cite{dezamarg86}, \cite{yoko94}, \cite{abraong93}.
In~\cite{dezamarg86} the intersection-type
theories which give $\l$-models where some filters are never
interpretations of $\l$-abstractions and which are complete for the
F-semantics are characterized.
More specifically it is shown that an intersection-type
theory $\sar{\tytv}$ satisfies the previous conditions if and only if
$\tyt\notteqt\tyt\to\tyt$, types
are invariant under
$\beta$-equality of subjects, and moreover the following rule (due to
R.Hindley):
$$\begin{array}[b]{lll}\label{hr}
 (\mbox{Hindley rule)}&\Ruled{\Gamma \ctas M:\psi\into(\tyt^{n}\to\tyt)
\;\;x_i\notin \FV(M)\;(1\leq i\leq n)}{\Gamma \ctas \lambda
    x_1\ldots x_n.Mx_1\ldots x_n:\psi} & \\[1em]
 \end{array}$$
for all $\psi\in\conv$,
is a derived rule.

Yokouchi in~\cite{yoko94} shows that if we add two suitable rules
(quite similar to Hindley rule) to the intersection-type assignment system
of~\cite{coppdezavenn81} we obtain completeness for the F-semantics.

Abramsky and Ong, in \cite{abraong93}, prove the completeness of the theory $\Sigma^{\abo}$, with respect to the
F-semantics, over applicative structures with convergence.

We conclude the paper with three final remarks.

If $\sar{\tytv}$ is a natural type theory
which is adequate for the F-semantics, then
Hindley's rule
is admissible in ${\ti}^{\tytv}_\tyt$. This follows from the observation that
under the given conditions for all $n\geq 0$ and for all $\psi\in\con^\tytv$ we can find
$I, A_i^{(1)}, \ldots, A_i^{(n)}, B_i$ such that
$$\psi\into(\tyt^{n}\to\tyt)\teqt\binto_{i\in I}(A_i^{(1)}\to
\ldots\to A_i^{(n)}\to B_i).$$
\newcommand{\1}{{\bf 1}}
We could not have used the syntactic approach based on term models
introduced in \cite{hind82} for showing all our adequacy results concerning the simple semantics , 
but not as far as the inference or the F-semantics. To this end, notice that if $\rho(x)=[M]$ and $M\I$ reduces 
to
an abstraction then a fortiori $M\1$ reduces to an abstraction, where $\I\equiv\l x.x$ and $\1\equiv\l xy.xy$.
Therefore
$\interpretation{x\1}{\rho}{\;}$ is the representative of a function
whenever $\interpretation{x\I}{\rho}{\;}$ is the representative of a
function. Now consider the F-type theory $\Sigma(\{\phi,\Omega\},{\cal DM})$
where ${\cal
DM}=\abo\cup\{\phi\cap(\Omega\to\Omega)\sim\Omega\to\phi\}$~\cite{dezamarg86}.
We have $x\:(\phi\to\phi)\to\Omega\to\Omega\vdash^{{\cal DM}}_\tyt
x\I:\Omega\to\Omega$ which implies, by soundness,
$x\:(\phi\to\phi)\to\Omega\to\Omega\ag^{{\cal DM}}_F
x\I:\Omega\to\Omega$. By the above we get
$x\:(\phi\to\phi)\to\Omega\to\Omega\ag^{{\cal DM}}_F
x\1:\Omega\to\Omega$, but it is easy to check, using the Generation
Lemma, that we cannot deduce $ x\1:\Omega\to\Omega$ from
$x\:(\phi\to\phi)\to\Omega\to\Omega$. As a matter of fact, the proof of completeness for the F-semantics
in~\cite{yoko94} uses a clever variant of the term model for a $\l$-calculus
with constants. It is not clear to us if this could be adapted to
the general case treated here.

It would be nice to investigate independent set-theoretic conditions
which imply that a type interpretation and a type environment agree
with a type theory. The canonical example in this sense is the one
given by partial applicative structures and the theory $\ehr$.

\cite{alesdezahons00} is an extended abstract of the present paper.\\

\noindent{\bf Acknowledgments}
The authors are grateful to Yoko Motohama and to the referees of ITRS submission for their useful
comments and suggestions.

\bibliography{refbook-dhm00}

\label{@lastpg}
\end{document}